\documentclass[aps,pra,twocolumn,showpacs,superscriptaddress,10pt]{revtex4-1}
\usepackage{bbm}
\usepackage{graphicx}
\usepackage{amsmath}
\usepackage{amssymb,amsthm}
\usepackage{hyperref}
\usepackage[normalem]{ulem}
\usepackage{xcolor}
\usepackage{mathtools}
\usepackage{yhmath}
\usepackage{mathbbol}
\usepackage{physics}
\usepackage{upgreek}

\begin{document}

\title{Neutral atom transport and transfer between optical tweezers}

\author{Cristina~Cicali}
\email[]{c.cicali@fz-juelich.de}
\affiliation{Forschungszentrum Jülich GmbH, Peter Grünberg Institute, Quantum Control (PGI-8), 54245 Jülich, Germany}
\affiliation{Institute for Theoretical Physics, University of Cologne, Zülpicher Straße 77, 50937 Cologne, Germany}

\author{Martino~Calzavara}
\affiliation{Forschungszentrum Jülich GmbH, Peter Grünberg Institute, Quantum Control (PGI-8), 54245 Jülich, Germany}
\affiliation{Institute for Theoretical Physics, University of Cologne, Zülpicher Straße 77, 50937 Cologne, Germany}

\author{Eloisa~Cuestas}
\affiliation{Forschungszentrum Jülich GmbH, Peter Grünberg Institute, Quantum Control (PGI-8), 54245 Jülich, Germany}
\affiliation{OIST Graduate University, Onna, Okinawa, Japan}

\author{Tommaso~Calarco}
\affiliation{Forschungszentrum Jülich GmbH, Peter Grünberg Institute, Quantum Control (PGI-8), 54245 Jülich, Germany}
\affiliation{Institute for Theoretical Physics, University of Cologne, Zülpicher Straße 77, 50937 Cologne, Germany}
\affiliation{Dipartimento di Fisica e Astronomia, Università di Bologna, 40127 Bologna, Italy}

\author{Robert~Zeier}
\affiliation{Forschungszentrum Jülich GmbH, Peter Grünberg Institute, Quantum Control (PGI-8), 54245 Jülich, Germany}

\author{Felix~Motzoi}
\email[]{f.motzoi@fz-juelich.de}
\affiliation{Forschungszentrum Jülich GmbH, Peter Grünberg Institute, Quantum Control (PGI-8), 54245 Jülich, Germany}
\affiliation{Institute for Theoretical Physics, University of Cologne, Zülpicher Straße 77, 50937 Cologne, Germany}

\date{\today}

\begin{abstract}
We focus on the optimization of neutral atom transport and transfer between optical tweezers, both critical steps towards the implementation of quantum processors and simulators. We consider four different types of experimentally relevant pulses: piece-wise linear, piece-wise quadratic, minimum jerk, and a family of hybrid linear and minimum jerk ramps. We also develop a protocol using Shortcuts to Adiabaticity (STA) techniques that allows us to include the effects of static traps. By computing a measure of the error after transport and two measures of the heating for transient times, we provide a systematic characterization of the performance of all the considered pulses and show that our proposed STA protocol outperforms the experimentally inspired pulses. After pulse shape optimization we find a lower threshold for the total time of the protocol that is compatible with the limit below which the increase in the vibrational excitations exceeds half of the amount of states hosted by the moving tweezer. Since the obtained lower bound for the atom capturing or releasing stage is 9 times faster than the one reported in state-of-the-art experiments, we interpret our results as a wake-up call towards the importance of the inclusion and optimization of the transfer between tweezers, which may be the largest bottleneck to speed. For the two pulses having the best performance (minimum jerk and STA), we determine optimal regions in the experimentally accessible parameters to implement high fidelity transport pulses. Finally, our STA results prove that a modulation in the depth of the moving tweezer designed to counteract the effect of the static traps reduces errors and allows for shorter pulse duration. To motivate the use of our STA pulse in future experiments, we provide a simple analytical approximation for the tweezer position and depth controls.  
\end{abstract}

\maketitle

\section{Introduction}
\label{sec_intro}

Over the last ten years we have witnessed a sustained growth in the capabilities for quantum information processing and simulation, which has been possible thanks to significant progress in both the confinement and control of atomic arrays \cite{bluvstein2022quantum, Browaeys2020, Daley2022, Morgado2021, Gyger2024, Barredo2016, Endres2016}. In particular, the precise manipulation and transport of neutral atoms in optical tweezers and lattices have become critical components in the development of quantum processors and quantum simulators \cite{bluvstein2024logical, Kaufman2021}. In state-of-the-art neutral atom quantum processors the atoms are usually cooled at a temperature in the range of $10-100\,\mu\text{K}$ and then stochastically loaded (with a probability of about $0.5$ or half-filling) from a magneto-optical trap (MOT) into a set of space and time controlled Gaussian traps or optical lattices. In general, the experimental setup involves a spatial light modulator (SLM) to create the static trap array while the moving optical tweezers are controlled by acousto-optic deflectors (AOD) \cite{bluvstein2022quantum, bluvstein2024logical, Barredo2016}; other approaches for similar experiments rely only on tweezers generated and controlled via AODs \cite{liu_prx_2019, picard_prx_2024, zhang_qstiop_2022, Kaufman2021}. These tweezers allow for a precise rearrangement of stochastically loaded atoms, enable non-local connectivity, and eliminate the need to prepare a new ensemble of atoms after each measurement, thus enhancing experimental efficiency \cite{Tindall24}. 

The basic requirements for a quantum processor include initialization and storage of qubits in a quantum register, bringing qubits sufficiently close to realize quantum gates and the final readout \cite{beugnon2007}. Therefore, efficient atom transport is not only a key step to perform controlled translations from the preparation or cooling chamber to the science cell; it also allows for on-demand interactions in order to realize quantum operations in the correct location (both processing and storing sites) and with the correct timing. In this context, fast and accurate manipulation of atomic motion emerges as a central need for quantum technologies in order to preserve coherences while attaining high fidelities between the obtained final state and a predefined target state \cite{Torrontegui2011}. Moreover, atom transport appears as a potential longterm bottleneck in quantum computing with cold atoms, given that other operations can potentially be significantly faster \cite{Chew2022}. The ideal goal of single-atom transport is to obtain a final state as close as possible to the initial state while avoiding losses and vibrational excitations. Though long transport times (adiabatic processes) may be the easiest solution to achieve high final fidelities, they also translate into the accumulation of decoherence and experimental noise \cite{Lam2021}. On the other hand, a diabatic (fast) process by default leads to higher excitations and thus to overheating and losses. Therefore, it is necessary to find a trade-off between high fidelity and transport time \cite{Endres2016, Barredo2016}. Our goal is to address this need using tools provided by quantum optimal control.

In the present work we focus on the optimization of atom transport and transfer between optical tweezers in order to speed up these operations while avoiding unwanted excitations. For that purpose we use optimal quantum control techniques to drive the system towards the desired state by minimizing the infidelity \cite{Lam2021, caneva2009optimal, zhang_pra_2015}, a measure of the deviation between the target state and the state obtained after the evolution of a particle moving from one optical trap to the next one. Even though the obtained evolution path should by construction lead to a final state close to the desired one there is no guarantee on the no-heating or no-loss condition, meaning that the obtained evolution might excite the system into the upper levels of the trap and therefore lead to atom losses during transport \footnote{Even though there is no dissipative mechanism considered in our model we adhere to the identification of vibrational excitations and heating, see for instance Refs.~\cite{gardiner_pra_2000, gehm_pra_1998, savard_pra_1997}}. To solve this, we firstly analyze and characterize the performance of several tweezer trajectories used in state-of-the-art experiments (we consider four families: linear, minimum jerk, quadratic, and a hybrid between the linear and minimum jerk ones \cite{schymik_PRA_2020, Endres2016, bluvstein2022quantum, bluvstein2024logical, liu_prx_2019, picard_prx_2024, zhang_qstiop_2022}) with particular focus on a measure of the error after transport and two measures of the vibrational excitations during transport as a function of the total time of the protocol. 

Besides considering experimentally motivated tweezers trajectories, we further develop a Shortcuts-to-Adiabaticity (STA) solution to generate analytical pulses, which can also serve as a seed for subsequent optimization. In general, the STA method leverages a Hamiltonian invariant to derive pulses that drive the system in a non-adiabatic (fast) way to the same final state as their adiabatic (slow) counterparts \cite{Torrontegui2011, LR1969}. The STA method provides great flexibility for computing the tweezer trajectory because it depends on ansatz functions that can be chosen freely as long as they satisfy the necessary boundary conditions. When addressing the atom transport problem, a wide variety of STA solutions have been derived for a harmonic trap or power-laws potential with time-independent frequency \cite{Torrontegui2011, Chen2011, zhang_pra_2015}. In our approach, we include the static tweezers potential in the harmonic approximation to obtain solutions tailored for our specific problem. We show that our STA protocol outperforms the best experimentally inspired transport pulse, and we provide analytical approximation formulas for our proposed STA pulses that can be straightforwardly implemented in current experiments. 

Following the characterization of the initial pulses we carry out optimal pulse shaping using the d-CRAB optimization algorithm \cite{caneva2011chopped, rach2015dressing, muller2022one} implemented in the user-friendly QuOCS toolkit \cite{rossignolo2023quocs}. This algorithm takes advantage of an expansion of the pulse over a randomized function basis that allows for a drastic reduction of the number of free parameters to be optimized, so that a direct search method (such as Nelder-Mead) can be employed. At the same time, the need for computing the gradient of the control objective (or figure of merit) is alleviated. This combined with the small number of optimization parameters with low bandwith, makes d-CRAB especially well-suited for direct experimental application. With this approach, low infidelities with the target state were achieved, improving by more than one order of magnitude compared to the analytical pulse shapes. All the optimized pulses present a more stable behavior of the fidelity with the target state after transport. Notably, while the optimized pulses reach an error threshold of $10^{-4}$ in a shorter total time compared to the non-optimized experimentally-motivated pulses (with reductions in time of 10-30\%), our proposed STA pulse requires the shortest time in order to obtain an error below $10^{-4}$ and that time does not change after optimization. We interpret this as a signature of the high quality and suitability of our STA solution, coming close to the numerically evaluated Quantum Speed Limit (QSL) \cite{deffner_jpa_2017,Lam2021}. By computing two measures of the vibrational excitations during transport, we rule out one of the four families of experimentally inspired pulses: the piece-wise linear one. For a given upper bound for the excitations (for instance, depending on the experimentally achievable trap depth) our results allow for the determination of a lower bound for the total protocol time.    

With the goal of identifying good regions in the pulses' parameter space to be used in the design of realistic experimental protocols, we choose the two pulses with the best performance and provide a heat map of the error after transport as a function of the total time and the depth of the moving tweezer. We determine a region in the parameter space for the minimum jerk pulse leading to errors below $10^{-4}$ and certain `magical time windows' for the STA pulse where the error is supressed by at least one order of magnitude. In the parameters interval we investigate, we observe that the performance of our STA pulse is almost independent of the moving tweezer depth. 

Finally, we identify a constraint in the choice of the total transport duration that is linked to the system's fundamental QSL \cite{deffner_jpa_2017,Lam2021}. For an experimentally achievable moving trap depth of about $ 3.57 \times 2\pi \text{ MHz}$ and a transport distance of $7 \text{ } \mu \text{m}$, our results point towards a quantum speed limit of about $ 8 \,\tau_{st}$ (here, for $^{39}$K atoms, $\tau_{st}\approx 0.03 \text{ ms}$  is the characteristic time of the static tweezers, related to the oscillator time during transport $\tau_{mt}$ via $\tau_{st} \approx 2\, \tau_{mt}$ for the considered maximum tweezer depth), a value that coincides with the time at which we observe an increase in the vibrational states equal to half of the states hosted by the trap. Even though our obtained time threshold for the transport process is roughly five times larger than the one reported in Ref.~\cite{pagano_prr_2024}, our results show a lower time threshold for the atom capture or release stage that is 9 times smaller than the one reported in the experiments of Ref.~\cite{spence_njp_2022}. Since the time required for capturing or releasing the atom was estimated to be 12 times larger than the time needed for the pure transport process \cite{schymik_PRA_2020}, and the protocol of Ref.~\cite{pagano_prr_2024} does not include this time cost, we interpret our results as a warning sign on the importance of considering the transfer between tweezers in the model. 

The paper is structured as follows: in section \ref{sec_model} we present our model for the atom transport problem. In section \ref{sec_pulses} we describe the considered experimental pulses together with the method to generate analytical pulses by means of a Shortcuts to Adiabaticity procedure. In section \ref{sec_transport_opt} we characterize the transport process for all the considered pulses and report the optimization results obtained using realistic experimental parameters. A summary and conclusions are given in section \ref{sec_concl}. Appendices \ref{sec_app_sta}-\ref{sec_app_transport_evol} contain details about the Shortcuts to Adiabaticity calculations, the numerical time evolution method, and the transport dynamics respectively.

\section{Atom transport in an external potential}
\label{sec_model}


The transport of an atom in the presence of a static external potential $V_{st}(x)$, from the initial position $x_i=0$ at time $t_i=0$ to the final position $x_f=d$ at time $t_f=T$ can be generically described by the Hamiltonian 
\begin{equation}
\label{eq:sim_01}
    H(x,t) =- \frac{\hbar}{2m}\frac{\partial^{2}}{\partial x^{2}} + V_{mt}(x,t) + V_{st}(x),
\end{equation}
where $m$ is the mass of the atom. In current experiments, the external potential $V_{st}(x)$ is typically generated either by a series of stationary optical tweezers leading to a sum of Gaussian traps, or by counter-propagating laser beams resulting in a sinusoidal pattern \cite{bloch2005ultracold, Yang2020, gross2017quantum, beugnon2007, saffman2010quantum, bluvstein2024logical, bluvstein2022quantum, Kaufman2021, chalopin2024optical, steinert2023spatially}. Here, we consider $^{39}$K atoms with $m = 6.47 \, 10^{-26} \text{ kg}$ and we focus on static Gaussian traps, meaning that the external potential $V_{st}(x)$ consists in a set of $n_{st}$ time-independent Gaussian wells defined by the depth amplitude $A_{st}/\hbar = 0.53 \times 2\pi \text{ MHz}$, the width $\sigma_{st}= 0.35 \, \mu$m, and the positions of the minima $x_{st}^{n} = (n-1)d$ with $d = 7\, \mu$m being the distance between two adjacent minima of the periodic potential and $n=1,2,..., n_{st}$ \cite{osterholz2020freely}. To model the moving tweezer we use a time dependent Gaussian potential $V_{mt}(x,t)$ and assume that the width of the tweezer is fixed to $\sigma_{mt}= 0.47 \,\mu\text{m}$ during the evolution \cite{osterholz2020freely, spence_njp_2022, schymik_PRA_2020,Tuebingen}, in which case the control parameters for our problem are the amplitude of the moving tweezer $A_{mt}(t)$ (trap depth) and the position of the center of the Gaussian well $x_{mt}(t)$. The static and moving potentials are respectively 
\begin{subequations}
\label{eq_potentials}
\begin{align}
\label{eq:sim_02a}
&V_{st}(x) = -A_{st}\sum_{n=1}^{n_{st}}\exp{\left(-\frac{(x-x_{st}^{n})^2}{2\sigma_{st}^2}\right)} \, , \text{ and} \\
\label{eq:sim_02b}
&V_{mt}(x,t) = -A_{mt}(t)\exp{\left(-\frac{(x-x_{mt}(t))^2}{2\sigma_{mt}^2}\right)}.
\end{align}
\end{subequations}
For the particle to be correctly transferred from the initial static tweezer to the moving one, transported through the total distance $d$, and then transferred to the final static tweezer, the following boundary conditions need to satisfied;
\begin{equation}
\begin{split}
    &A_{mt}(t_i)=0 \text{ } \text{and} \text{ } A_{mt}(t_f)=0, \\
    &x_{mt}(t_i)= x_i= 0 \text{ } \text{and} \text{ } x_{mt}(t_f)=x_f=d \, .
\end{split}
\label{eq:sim_03}
\end{equation}
While the maximum amplitude for both the moving tweezer and the static traps are given by the power of the laser beams, the dynamics of the transport protocol depends on the trade-off between the static and moving amplitudes. Comparable amplitudes translate into a strong interaction between the moving and static tweezers with the consequent oscillations of the atom in the traps that, in turn, lead to losses during transport. To prevent this from happening, we impose an extra condition on the amplitude during transport: $A_{mt}>A_{st}$. 

\begin{figure}[tb]
    \centering
    \includegraphics[width=0.48\textwidth]{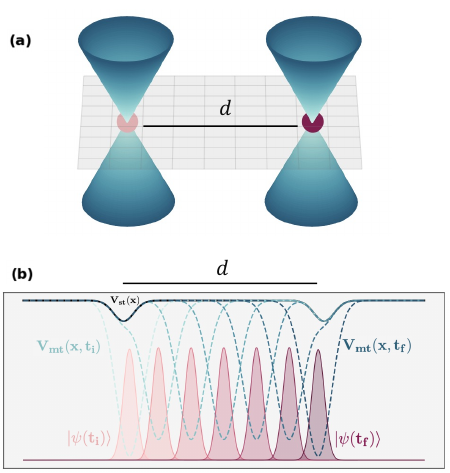}
    \caption{Illustration of the transport protocol. (a) Two adjacent static tweezers (blue) separated by a distance $d$. (b) The atom initially trapped at $x_i=0$ is transported to the position $x_f=d$ in a total time $T=t_f-t_i$. The amplitude of the external (static tweezer) potential denoted by $V_{st}(x)$ is shown in full solid black line while the one of the moving tweezer $V_{mt}(x,t)$ is depicted as dashed blue lines with a color gradient indicating the time evolution (from $t_i$ to $t_f$ lighter to darker -- left to right). The probability of finding the particle in the position $x$ is given by $\abs{\psi(t)}^2$ and it is represented by the filled curves with a purple-color gradient associated as before with the time arrow.}
    \label{fig:ts_01}
\end{figure}

Figure (\ref{fig:ts_01}) presents a scheme of the atomic transport protocol. The initial state $\ket{\psi_i}$, and final target state $\ket{\psi_{tg}}$, are the ground states (computed via exact diagonalization) of the static Hamiltonian with a single Gaussian potential centered respectively at $x_i=0$ and $x_f=d$. The goal is to determine the controls $x_{mt}(t)$ and $A_{mt}(t)$ such that the state after the complete time evolution, denoted by $\ket{\psi(t_f)}$, closely approximates the localized ground state of the final targeted trap $\ket{\psi_{tg}}$. To quantify the error after the transport protocol we compute the infidelity $\mathcal{I}=1-\mathcal{F}$, where $\mathcal{F}$ is the fidelity defined as the overlap $\mathcal{F} = \abs{\braket{\psi(t_f)}{\psi_{tg}}}^{2}$. In order to shape the control pulses that execute the transport task with low infidelity, we first characterize the performance of several different pulse shapes based on current experimental realizations together with pulses that we derive within the Shortcuts to Adiabaticity (STA) formalism. In a second stage, we use these pulse profiles as initial guesses for optimal pulse shaping using the d-CRAB algorithm, which allows to reach even higher and more stable fidelities. 


\section{Transport pulses}
\label{sec_pulses}

This section presents the theoretical description of the considered pulses. In Sec.~\ref{sec_pulses_desc} we describe the initial guesses motivated by experimental considerations. Section \ref{sec_pulses_sta} contains the analytical derivation of the Shortcuts to Adiabaticity (STA) solution. 

\subsection{Initial guesses based on experiments}
\label{sec_pulses_desc}

In current experiments the choice of the ramps that control the position of the moving tweezer aims to suppress heating and losses, or equivalently, to minimize excitations during and after transport \cite{Endres2016, bluvstein2022quantum, bluvstein2024logical, liu_prx_2019, schymik_PRA_2020, picard_prx_2024, zhang_qstiop_2022}. In our model the control pulses time-dependently define the spatial trajectory followed by the bottom of the Gaussian trap $x_{mt}(t)$ and its depth, determined by the amplitude $A_{mt}(t)$. In practice the minimum of the tweezer potential coincides with the focal spot \cite{pesce_2020_EPJP}, while the amplitude is proportional to the power of the beam \cite{simon_book_2016}. The four transport trajectories based on current experiments that we consider are: piece-wise linear, piece-wise quadratic, minimum jerk (derivative of the acceleration), and a hybrid trajectory that combines the linear and the minimum jerk ones. These are defined below.

The linear pulse, used for instance in Ref.~\cite{schymik_PRA_2020}, has as the advantage of simplicity and the fact that the constant velocity can be implemented easily in the experiments (constant sweep rate for the AOD frequency), but it is known to cause excitation due to the abrupt change in the transport velocity at the discontinuity points. In the case of the piece-wise quadratic trajectory used in the experiments of Ref.~\cite{Endres2016}, the atoms experience alternated accelerations of the same magnitude in the first and second half of the trajectory. This kind of quadratic pulse can be easily implement with frequency control systems \cite{matthies_thesis_2023}, however, the discontinuity in the acceleration might induce heating. The need for smooth position functions arises then to avoid unwanted excitations or energy excess. While introduced in a completely different context (voluntary movements on primates \cite{hogan_1984}) 40 years ago, the minimum jerk trajectory satisfies the latter requirement and was recently used to study diatomic molecule formation with the associated requirement of having not only the two atoms close enough but also in their relative motional ground state \cite{liu_prx_2019, picard_prx_2024, spence_njp_2022}. This trajectory minimizes the square of the jerk (derivative of the acceleration) over the full path, and it is also obtained when a polynomial ansatz is used to solve the position as a function of time for a translation with zero initial and final velocity and acceleration. In Ref.~\cite{liu_prx_2019} the authors argue that the selection of the trajectory should aim to minimize heating due to jerk at end points, and to avoid parametric heating due to trap depth oscillations \cite{jauregui_pra_2001}. The latter condition is further explained in Ref.~\cite{spence_njp_2022}; when the frequency of the AOD is driven with a constant sweep rate (associated with a linear translation), it helps to avoid resonant intensity modulations arising from imperfections in the AOD (see also Ref.~\cite{zhang_qstiop_2022}). A good compromise between preventing heating from changes in the acceleration at the beginning and end of the movement and at the same time avoiding resonant intensity modulations is found by using a hybrid pulse that implements a minimum jerk trajectory for the start and end of the ramp and a constant-velocity (linear) translation for intermediate times. 

As polynomial or piece-wise polynomial functions of time, the linear, quadratic, minimum jerk, and hybrid trajectories have a simple analytical form. Since the target state in the final tweezer should match closely the initial state in the initial tweezer, and the considered static tweezers only differ in their position, one can chose symmetric ``palindromic" pulses to reduce the degrees of freedom. In particular, the four families of experimentally motivated pulses that we consider are time-reversal symmetric. The linear trajectory $x^{l}$ is given by
\begin{equation}
\label{x_tr_linear}
x^l(d, \uptau, t) = d \, \frac{t}{\uptau} \, ,
\end{equation}
where $d$ is the covered distance and $\uptau$ is the total time required for transport. 

The piece-wise quadratic trajectory $x^{q}$, with a position described by two parabolas that intersect in the mid-point and have piece-wise constant acceleration, can be written as
\begin{equation}
\label{x_tr_pw_quad}
x^q(d, \uptau, t) = d 
    \begin{cases}
    \phantom{-} 2 \left(\frac{t}{\uptau} \right)^2  & \text{ for } 0 \leq t \leq \frac{\uptau}{2} \\
    - 2 \left(\frac{t}{\uptau} \right)^2 + 4 \, \frac{t}{\uptau} - 1  & \text{ for } \frac{\uptau}{2}  \leq t \leq \uptau \, .
    \end{cases}
\end{equation}

In turn, the minimum jerk trajectory $x^{mj}$ reads
\begin{equation}
\label{x_tr_min_jerk}
x^{mj}(d, \uptau, t) = d \left( 10 \left(\frac{t}{\uptau} \right)^3 - 15 \left(\frac{t}{\uptau} \right)^4 + 6 \left(\frac{t}{\uptau} \right)^5 \right) \, ,
\end{equation}
where it is straightforward to check that the initial ($t=0$) and final ($t=\uptau$) velocity and acceleration are equal to zero. 

Finally, the hybrid trajectory $x^{hyb}$ is defined upon the fraction $\upxi$ of the total transport time that follows a linear motion. This fraction parameter is also called hybridicity \cite{spence_njp_2022}. Since the total time under linear motion $\upxi \uptau$ ranges between $0$ and $\uptau$, we have that $0 \leq \upxi \leq 1$. Moreover, for $\upxi =0 $ the trajectory reduces to the minimum jerk one while for $\upxi = 1$ the trajectory coincides with the piece-wise linear one, so it is reasonable to expect that when changing the parameter $\upxi$ the dynamics of the system should interpolate between the dynamics under the linear and minimum jerk pulses. The hybrid trajectory can be written as   
\begin{align}
\label{x_tr_hyb}
x^{hyb}(d, \uptau, t) = 
    \begin{cases}
     x^{mj}(d \frac{8(1-\upxi)}{8+7\upxi}, \uptau (1-\upxi), t) \\ 
      \;\;\;\;\;\;\;\;\;\;\;\;\;\;\;\;\;\;\;\;\;\;\;\; \text{ for } 0 \leq t \leq \frac{1-\upxi}{2} \uptau, \\ \\
     d \left( \frac{15}{8+7\upxi} \frac{t}{\uptau} - \frac{7(1-\upxi)}{2(8+7\upxi)} \right) \\  \;\;\;\;\;\;\;\;\;\;\;\;\;\;\;\;\;\;\;\;\;\;\;\; \text{ for } \frac{1-\upxi}{2} \uptau  \leq t \leq \frac{1+\upxi}{2} \uptau, \\ \\
     x^{mj}(d \frac{8(1-\upxi)}{8+7\upxi}, \uptau (1-\upxi),  t- \uptau \upxi) +
     d \frac{15 \upxi}{8+7\upxi}   \\\;\;\;\;\;\;\;\;\;\;\;\;\;\;\;\;\;\;\;\;\;\;\;\; \text{ for } \frac{1+\upxi}{2} \uptau  \leq t \leq \uptau \, . \\
    \end{cases} 
\end{align}
In Ref.~\cite{liu_prx_2019} the experimental values of $\upxi$ are $0$ and $0.95$, i.e.\ a full minimum jerk trajectory and a $95\%$ linear one. In Ref.~\cite{spence_njp_2022} the authors use $\upxi = 0.1$ [see their figure 3(b)], value with which they report to avoid resonant intensity modulations for a translation of $4.5 \,\mu\text{m}$ and $\uptau \approx 1.3 \text{ ms}$. 

\begin{figure}[htb]
    \centering
    \includegraphics[width=0.45\textwidth]{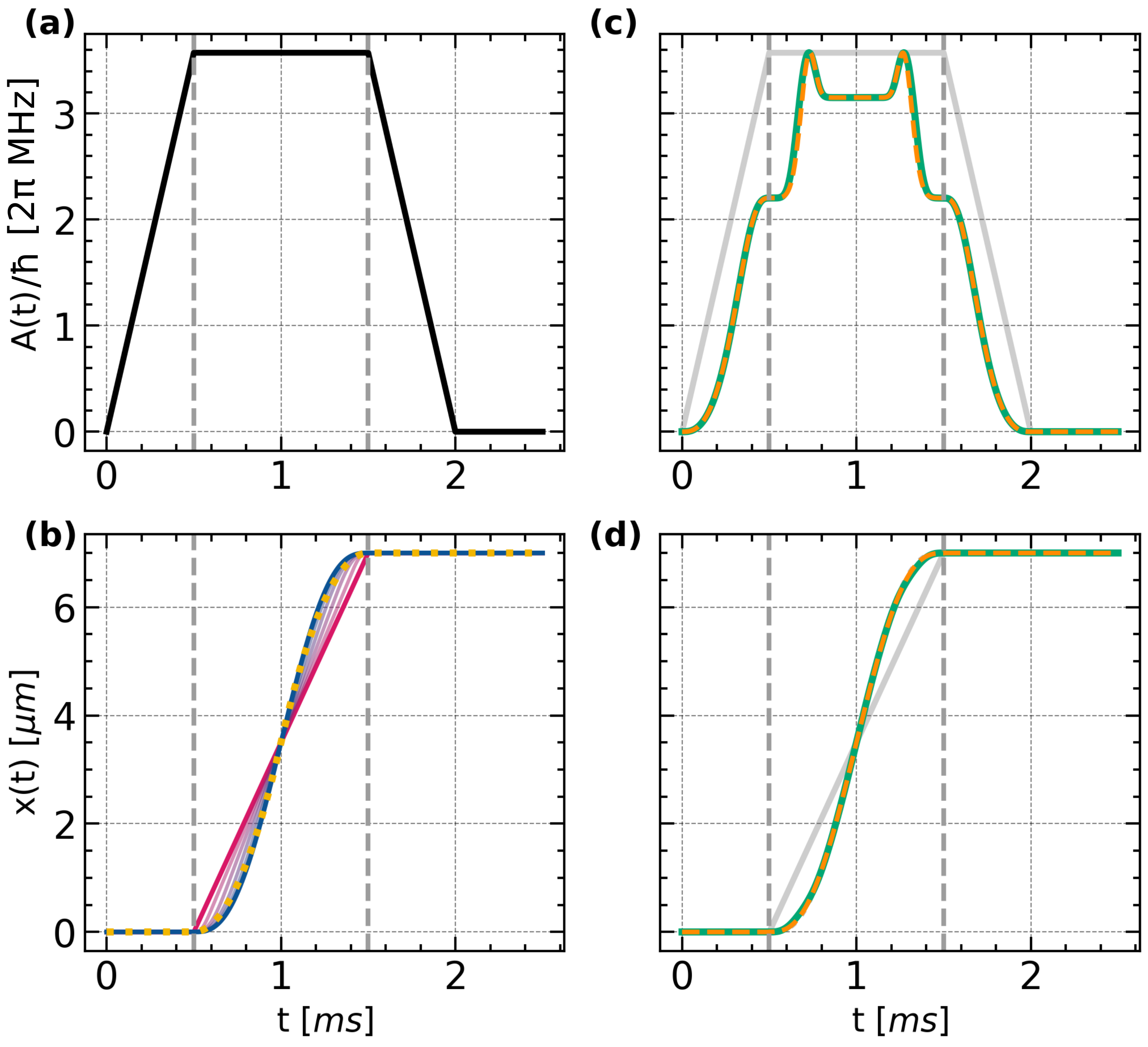}
    \caption{Position $x_{mt}(t)$ and amplitude $A_{mt}(t)$ control pulses. The depth of the moving tweezer or amplitude control is shown in panels (a) and (c), while the position of the minima of the trap is shown in panels (b) and (d) for a total time of $T=2.5 \text{ ms}$, a maximum amplitude $A_{mt}^{max}/\hbar= 3.57 \times 2\pi \text{ MHz}$, a transport time of $\eta \, T$ with $\eta=2/5$, and a total covered distance between static tweezers $d = 7\, \mu\text{m}$.  The initial and final transport times are indicated by gray dashed vertical lines. (a) Piece-wise amplitude ramp consisting of four stages: transferring the atom from the initial static tweezer to the moving one, transport, transferring the atom from the moving tweezer to the target static one, and a final waiting or measuring time [see Eq.~\eqref{A_mt_of_t}] (b) Position of the moving tweezer as a function of time for a piece-wise linear (magenta line), minimum jerk (blue line), and quadratic (yellow dashed line) pulse, see Eqs.~\eqref{x_tr_linear}-\eqref{x_tr_min_jerk}. The linear and minimum jerk trajectories are particular cases of the more general hybrid pulse given in Eq.~\eqref{x_tr_hyb}, which is also depicted for hybridizations $\upxi = 0.2,\,0.4,\,0.8$ (curves with a color gradient from blue to magenta). The velocity of the moving tweezer at the beginning and end of the transport interval decreases with $\upxi$ and is minimum for the minimum jerk pulse. (c) Amplitude shape calculated within the STA approach, full numerical solution (solid turquoise curve) and the approximation obtained using Eq.~\eqref{eq_x_mt_approx} (dashed orange line). The piece-wise linear amplitude associated to the piece-wise linear, hybrid, minimum jerk, and quadratic ramps is shown as a light gray curve to highlight that the maximum amplitude is the same for all pulses. The two peaks at maximum amplitude depicted by the STA pulse are developed to counteract the restoring force of the static tweezers. (d) Full numerical and approximated position of the moving tweezer obtained with the STA formalism, same color code as in panel (c).}
    \label{fig:ts_02}
\end{figure}

As mentioned before, we account not only for the transport process but also the transfer from the initial static trap to the moving tweezer and from the moving tweezer towards the target static one. Therefore, the protocol consists of three main stages: capturing the particle initially in the first static trap, the transport, and the final release of the particle in the target static tweezer. In the capturing stage, the depth of the moving tweezer is monotonically raised from zero to its maximum amplitude $A_{mt}^{max}$ while the position is kept constant. In the second transport stage, the position of the focal point follows one of the trajectories of Eqs.~\eqref{x_tr_linear}-\eqref{x_tr_hyb}, while the power of the laser is kept constant, consistently with the experiments described in \cite{picard_prx_2024}. The third releasing stage is the inverse process of capturing: while maintaining the target position, the depth of the moving tweezer is decreased to zero. To asses the dynamics and quality of the final state we add a fourth stage which we call waiting time. During this time the moving tweezer is off and we evaluate several statistical measurements over the final state to examine mainly its time-stability and the possible presence of oscillations after transport (see e.g. \cite{pagano_prr_2024}). Moreover, this waiting time can be used in the experiments to recool the atom before, for instance, the next concatenated transport step. The amplitude that we consider is then given by
\begin{equation}
\label{A_mt_of_t}
A_{mt}(t) = A_{mt}^{max} 
    \begin{cases}
    \frac{3}{1-\eta} \frac{t}{T}  & \text{ for } 0 \leq t \leq \frac{1-\eta}{3} T \\
    1  & \text{ for } \frac{1-\eta}{3} T  \leq t \leq \frac{1+2\eta}{3} T \\
    \frac{2+\eta}{1-\eta} - \frac{3}{1-\eta} \frac{t}{T} & \text{ for } \frac{1+2\eta}{3} T  \leq t \leq \frac{2+\eta}{3} T \\
    0 & \text{ for } \frac{2+\eta}{3} T  \leq t \leq T \, , \\ 
    \end{cases}
\end{equation}
where we have taken $t_i =0$ and $t_f =T$ and we divided the total time $T$ into fractions defined by the parameter $\eta$; the transport time is assigned to be $\eta \, T$ and the remaining time is divided into three intervals of equal lenght $(1-\eta)T/3$ for the capturing, releasing, and waiting stages. By doing so, we respect the symmetry condition between the capturing and releasing stage and we allocate a waiting time of length comparable to the other stages. The obtained total position pulse is
\begin{equation}
\label{x_mt_of_t}
x_{mt}(t) = x_i + 
    \begin{cases}
    0  & \text{ for } 0 \leq t \leq \frac{1-\eta}{3} T \\
    x(d , \eta \, T, t-\frac{1-\eta}{3} T)  & \text{ for } \frac{1-\eta}{3} T  \leq t \leq \frac{1+2\eta}{3} T \\
    d & \text{ for } \frac{1+2\eta}{3} T  \leq t \leq \frac{2+\eta}{3} T \\
    d & \text{ for } \frac{2+\eta}{3} T  \leq t \leq T \, , \\
    \end{cases}
\end{equation}
with $x(d,\uptau,t)$ being any of the trajectories defined in Eqs.~\eqref{x_tr_linear}-\eqref{x_tr_hyb}.

The respective pulses are shown in Figure \ref{fig:ts_02} (a) and (b). Panel (a) shows the amplitude obtained via Eq.~\eqref{A_mt_of_t} for a maximum amplitude $A_{mt}^{max}/\hbar = A_{exp}/\hbar = 3.57 \times 2\pi \text{ MHz}$, $d = 7\, \mu\text{m}$, $\eta = 2/5$, and $T=2.5 \text{ ms}$ \cite{Tuebingen}. The starting and ending time of the transport stage are indicated as gray dashed vertical lines. Both controls $A_{mt}(t)$ and $x_{mt}(t)$ satisfy the boundary conditions specified by Eq.~\eqref{eq:sim_03}. While the amplitude for the experimentally motivated pulses is always a piece-wise linear function, all the trajectories have an s-shape position dependence. In panel (b) the piece-wise linear (solid magenta) and minimum jerk trajectory (solid blue) are presented as the limiting cases of the hybrid trajectory, which is depicted for several $\upxi$ values (gradient from magenta to purple). The minimum jerk trajectory has the smallest change ratio at the beginning and at the end of the transport and reaches its highest velocity in the mid-time point. For the hybrid pulse the change ratio at the beginning and at the end of the transport decreases for decreasing hybridicity $\upxi$. The quadratic pulse (dashed yellow line) has a higher velocity in the first and final part of the transport when compared to the minimum jerk one, and it also has a slightly higher velocity in the mid-time or inflection point. 


 
\subsection{Initial guess based on Shortcuts To Adiabaticity}
\label{sec_pulses_sta}

Since the piece-wise linear, hybrid, minimum jerk, and quadratic control pulses are experimentally inspired ansätze for transport, they do not include the specifics of the system which are encoded in the Hamiltonian of Eq.~\eqref{eq:sim_01}. In particular, there is no consideration for the shape of the moving tweezers $V_{mt}(x,t)$ let alone the presence of the static tweezers $V_{st}(x)$ in the background [see Eq.~\eqref{eq_potentials}]. To derive pulses that solve the fast transport problem taking into account these effects while offering flexibility for satisfying desirable and realistic boundary conditions, we draw upon the shortcuts to adiabaticity (STA) framework. The main idea of STA methods is to develop protocols that lead to the same final state as their adiabatic (slow) counterpart in considerably shorter times \cite{guery_odelin_rmp_2019}, enabling us to avoid noise and decoherence effects. 

The minimum jerk trajectory was proposed to perform a translation of the minimum point of a perfect harmonic oscillator with minimal motional excitation at the beginning and end of the transport by constraining the velocity and acceleration to be zero at the initial and final times (see for instance Ref.~\cite{Torrontegui2011}). Here we use the minimum jerk functional form in intermediate steps of the derivation of a new set of controls $A_{mt}(t)$ and $x_{mt}(t)$ which take into account the specific form of the moving and static tweezers. While several STA-based approaches to atom transport have been proposed \cite{Torrontegui2011, Ness2018, Chen2011, zhang_pra_2015, Zhang2016, hauck_prapp_2022, guery_odelin_pra_2014}, to the best of our knowledge, the background of static tweezers has not previously been incorporated in an STA solution. In order to address a more realistic scenario, we derive an STA control pulse that includes the transfer between optical tweezers in the transport, therefore avoiding errors during this critical stage. 

The first step in our approach is to approximate the Hamiltonian of Eq.~\eqref{eq:sim_01} with a time-dependent harmonic oscillator and then use known results on fast eigenstate transfer for the harmonic case \cite{Dhara1984, Torrontegui2011}. In this way, the time dependent oscillator accounts for both the static and moving tweezers. We start by expanding the complete potential in a Taylor series up to the second order around the center of the moving tweezer $x_{mt}(t)$. By completing the resulting polynomial in $x$ to a square form and ignoring a time-dependent energy offset, we obtain a moving harmonic oscillator with a time dependent effective frequency
\begin{equation}
H_0(x,t) = - \frac{\hbar}{2m}\frac{\partial^{2}}{\partial x^{2}} + \frac{m\omega(t)^2}{2}[x - x_0(t)]^2 \,, 
\label{eq:H_harm}
\end{equation}
where $\omega(t)$ and $x_0(t)$ are related to the controls and the derivatives of the static potential through
\begin{equation}
m\omega^2(t) = \left.\frac{d^2V_{st}}{dx^2}\right\vert_{x_{mt}(t)} + \frac{A_{mt}(t)}{\sigma_{mt}^2} \, , 
\label{eq:amp}
\end{equation}
and
\begin{equation}
    x_0(t) = x_{mt}(t) - \frac{\left.\frac{dV_{st}}{dx}\right\vert_{x_{mt}(t)}}{m\omega^2(t)} \, . 
\label{eq:pos}
\end{equation}
For the time dependent Hamiltonian of Eq.~\eqref{eq:H_harm}, it is possible to construct a dynamical invariant $I(x,t)$ such that for any wave function $\psi(t)$ evolving with $H_0(x,t)$ we have $\dv{t} \expval{I(t)}{\psi(t)} =0 $. This means that with the appropriate time dependence of $\omega(t)$ and $x_0(t)$, if the system is initially in an eigenstate of $H_0(t_i)$, the time evolution will map it onto the corresponding eigenstate of $H_0(t_f)$ - see Appendix \ref{sec_app_sta} for details. For our particular form of $H_0(x,t)$, the dynamical invariant involves two auxiliary functions $\alpha(t)$ and $\rho(t)$ that satisfy
\begin{align}
    x_0(t) &= \frac{\ddot{\alpha}(t)}{\omega^2(t)} + \alpha(t) \,, \text{ and } \label{eq:q_0} \\
    \omega^2(t) &= \frac{\omega_0^2}{\rho^4(t)} - \frac{\ddot{\rho}(t)}{\rho(t)} \,, \label{eq:omga}
\end{align}
where $\omega_0$ is a free constant. From the first of these equations we see that $\alpha(t)$ can be identified with a classical trajectory of a driven generalized classical harmonic oscillator \cite{kumar_singh_optcomm_2010}, and $\rho(t)$ can be seen as a spatial rescaling factor. While Eqs.~\eqref{eq:q_0} and \eqref{eq:omga} ensure that $H_0(x,t)$ has fast transport modes given by the eigenstates of $I(t)$ up to a phase, we also need them to coincide with the eigenstates of $H_0(x,t)$ at the initial and the final time. To fulfill the latter requirement the two auxiliary functions must also satisfy the following boundary conditions, 
\begin{align}
    \dot{\alpha}(t_i) &= \ddot{\alpha}(t_i) = \dot{\alpha}(t_f) = \ddot{\alpha}(t_f) = 0 \,, \text{ and } \label{eq:bc_alpha}\\
    \dot{\rho}(t_i) &= \ddot{\rho}(t_i) = \dot{\rho}(t_f) = \ddot{\rho}(t_f) = 0 \,. \label{eq:bc_rho}
\end{align}
The strategy to obtain the desired pulse is essentially reverse engineering. Its starting point is to choose $\alpha(t)$ and $\rho(t)$ satisfying the boundary conditions, which we then use to compute $x_0(t)$ and $\omega(t)$ using Eqs.~\eqref{eq:q_0} and \eqref{eq:omga}. After that, we numerically solve Eq.~\eqref{eq:pos} to find the position of the moving tweezer $x_{mt}(t)$, which is finally plugged in Eq.~\eqref{eq:amp} to obtain the tweezer amplitude $A_{mt}(t)$. 

A simple way to design functions satisfying boundary conditions on the derivatives such as the ones of Eqs.~\eqref{eq:bc_alpha} and \eqref{eq:bc_rho} is to look for an appropriate polynomial function \cite{Ness2018, Torrontegui2011, Campo2012, Chen2010}. This polynomial interpolation has the advantage that most of the calculations can be carried out analytically. As already mentioned, the lowest order polynomial in a variable $s$ having vanishing first and second derivative at $s=0,1$ (initial and final points) is given by the minimum jerk polynomial $p_{mj}(s) = 10 s^3 - 15s^4 + 6s^5$ (notice that $p_{mj}(0) = 0$ and $p_{mj}(1) = 1$), therefore, we choose both $\alpha(t)$ and $\rho(t)$ to have the functional form of $p_{mj}$ with some multiplicative and additive constant. Because our transport protocol is divided in stages, we apply the reverse engineering process to each time interval and require continuous outputs. In other words, for each stage we propose $\alpha(s) = \alpha_i + p_{mj}(s) (\alpha_f-\alpha_i)$ and $\rho(s) = \rho_i + p_{mj}(s) (\rho_f-\rho_i)$ where $s$ is the adimensional time $s= t/(t_f - t_i)$ and the $i$ and $f$ subindices denote initial and final values over the considered interval. Using this form for the auxiliary functions, Eqs.~\eqref{eq:bc_alpha} and \ref{eq:bc_rho} are satisfied for any $\alpha_{i,f}$ and $\rho_{i,f}$, which are to be fixed by the initial and the target position and by requiring the physical controls to be continuous functions of time. 

Now we turn to the calculation of the auxiliary functions. As mentioned before, for the proposed form for $\alpha(t)$ it is easy to see that $\ddot{\alpha}(t)$ vanishes at the initial and the final time of each stage, i.e.\ for $t=0,\, (1-\eta)T/3,\, (1+2\eta)T/3,\, (2+\eta)T/3 ,\, T$, therefore for those time values we have $\alpha(t) = x_0(t)$. Since the position of the moving optical tweezer $x_0(t)$ only changes during the transport interval the calculation of $\alpha(t)$ is straightforward and leads to
\begin{widetext}
\begin{equation}
\label{eq_alpha_of_t}
\alpha(t) = d
    \begin{cases}
    0 & \text{ for } 0 \leq t \leq \frac{1-\eta}{3} T \\
    10 \left( \frac{t - \frac{(1-\eta)}{3} T}{\eta T} \right)^3 -15 \left( \frac{t - \frac{(1-\eta)}{3} T}{\eta T} \right)^4 +6 \left( \frac{t - \frac{(1-\eta)}{3} T}{\eta T} \right)^5   & \text{ for } \frac{1-\eta}{3} T  \leq t \leq \frac{1+2\eta}{3} T \\
    1 & \text{ for } \frac{1+2\eta}{3} T  \leq t \leq \frac{2+\eta}{3} T \\
    1 & \text{ for } \frac{2+\eta}{3} T  \leq t \leq T \, . \\
    \end{cases} 
\end{equation}
\end{widetext}
To compute the auxiliary function $\rho(t)$ it is useful to define the frequency associated to the static traps
\begin{equation}
\label{eq_omegast}
    \omega_{st} = \sqrt{\frac{1}{m}\left.\frac{d^2V_{st}}{dx^2}\right\vert_{x=0,\,d}} = \sqrt{\frac{A_{st}}{m\sigma_{st}^2}} \, ,
\end{equation}
and the maximum frequency for the moving tweezer over the capturing and releasing intervals, which is the same as the initial frequency in the transport interval and is given in terms of the maximum deepth of the moving tweezer over the capturing ($c$) or releasing ($r$) stages $A_{mt}^{max,cr}$
\begin{equation}
\label{eq_omegamtmax}
    \omega_{mt}^{max,cr} = \sqrt{\frac{A_{mt}^{max,cr}}{m \sigma_{mt}^2}} \, .
\end{equation}
Notice that in the case of the experimentally inspired pulses described in the previous section $A_{mt}^{max,cr}$ matches the global maximum amplitude of the pulse $A_{mt}^{max}$. At the beginning of the capturing stage ($t=0$), at the end of the releasing interval ($t=(2+\eta)T/3$), and at the end of the protocol ($t=T$) the moving tweezer is completely off and therefore we have that $\omega(t)= \omega_{st}$ [see Eq.~\eqref{eq:amp}]. 

On the other hand, at the end of the capturing and beginning of the releasing interval, $t=(1-\eta)T/3$ and $t=(1+2\eta)T/3$ respectively, the amplitude of the moving tweezer is at its maximum value over the capturing or releasing stage meaning that $\omega(t) = \sqrt{\omega_{st}^2+{(\omega_{mt}^{max,cr})}^2}$. By using this in Eq.~\eqref{eq:omga} we get
\begin{widetext}
\begin{equation}
\label{eq_rho_of_t}
\rho(t) = \frac{\sqrt{\omega_0}}{\sqrt{\omega_{st}}}
    \begin{cases}
    1 + \left( \frac{1}{\sqrt[4]{\tilde{\omega}^2+1}} -1  \right) \left\lbrace 10 \left( \frac{t}{\frac{(1-\eta)}{3} T} \right)^3 -15 \left( \frac{t}{\frac{(1-\eta)}{3} T} \right)^4 +6 \left( \frac{t}{\frac{(1-\eta)}{3} T} \right)^5  \right\rbrace  & \text{ for } 0 \leq t \leq \frac{1-\eta}{3} T \\
    \frac{1}{\sqrt[4]{\tilde{\omega}^2+1}}  & \text{ for } \frac{1-\eta}{3} T  \leq t \leq \frac{1+2\eta}{3} T \\
    \frac{1}{\sqrt[4]{\tilde{\omega}^2+1}} - \left( \frac{1}{\sqrt[4]{\tilde{\omega}^2+1}} -1  \right) \left\lbrace 10 \left( \frac{t - \frac{(1+2\eta)}{3} T}{\frac{(1-\eta)}{3} T} \right)^3 -15 \left( \frac{t -\frac{(1+2\eta)}{3} T}{\frac{(1-\eta)}{3} T} \right)^4 +6 \left( \frac{t -\frac{(1+2\eta)}{3} T}{\frac{(1-\eta)}{3} T} \right)^5  \right\rbrace & \text{ for } \frac{1+2\eta}{3} T  \leq t \leq \frac{2+\eta}{3} T \\
    1 & \text{ for } \frac{2+\eta}{3} T  \leq t \leq T \, , \\
    \end{cases} 
\end{equation}
\end{widetext}
with $\tilde{\omega}^2 = (\omega_{mt}^{max,cr}/\omega_{st})^2 = A_{mt}^{max,cr} \sigma_{st}^2/ (A_{st} \sigma_{mt}^2)$. Now that we have the expression of $\rho(t)$, the calculation of $\omega(t)$ is straightforward via Eq.~\eqref{eq:omga}. The resulting $\omega(t)$ can be used together with $\alpha(t)$ in Eq.~\eqref{eq:q_0} to obtain $x_0(t)$. It is important to notice that since $\rho(t) \propto \sqrt{\omega_0}$ and also because $\alpha(t)$ does not depend on this quantity, then $\omega(t)$ and as a consequence $x_0(t)$ are independent of $\omega_0$. After that, the obtained $x_0(t)$ is plugged in Eq.~\eqref{eq:pos} and the equation is numerically solved to get $x_{mt}(t)$, which we then use in Eq.~\eqref{eq:amp} in order to derive the amplitude control $A_{mt}(t)$. 

The numerical trajectories for the amplitude and the position of the bottom of the moving trap are shown in Fig. \ref{fig:ts_02}(c) and (d) as turquoise solid curves. To compare the obtained solution with the pulses described in the previous section, we superimpose the piece-wise linear pulse and the linear piece-wise amplitude as light gray solid curves. While the obtained $x_0(t)$ strongly resembles the minimum jerk trajectory, the amplitude presents two peaks that counteract the effect of the static tweezers during the atom transport task and are directly related to the second derivative term in Eq.~\eqref{eq:amp} -- see also Appendix \ref{sec_app_sta}. For a general problem, the quantity and spacing of these peaks will be given by both the amount of static traps and the spatial separation between them. In most of the experimental implementations, the power of the laser that generates the moving tweezer is kept constant during the movement (see for instance Ref.~\cite{picard_prx_2024} or figure 9 of Ref.~\cite{spence_njp_2022}), as we already consider for the piece-wise linear, hybrid, minimum jerk, and quadratic ramp. However, in the supplemental material of Ref.~\cite{bluvstein2024logical} the authors mention that when transferring from the static to the moving tweezers they use a quadratic intensity profile, highlighting the possibility to implement our STA amplitude pulse. 

Before moving to the next section, we provide a quite simple and directly applicable analytical approximation for the STA controls. The approximated solutions can be obtained by using a quadratic approximation for the external potential, $V_{st}(x) \approx - A_{st} (1-2 x/d)^2$, only valid between the two minima, i.e.\ for $0 \leq x \leq d$. Within this approximation we get
\begin{equation}
\label{eq_x_mt_approx}
x_{mt}(t) \approx  \frac{x_0(t)+ \frac{4 A_{st}}{d m \omega^2(t)}}{1+\frac{8 A_{st}}{d^2 m \omega^2(t)}} \, ,
\end{equation}
an expression that when combined with Eq.~\eqref{eq:amp} leads to an approximated solution for $A_{mt}(t)$. It is important to notice that, for this approximated $x_{mt}(t)$, the boundary conditions are also approximately fulfilled. In the transport interval, the approximated solution can be written as
\begin{widetext}
\begin{equation}
\label{eq_x_mt_approx_2}
x_{mt}(t) \approx d \frac{ 10 \left(\frac{t - \frac{1-\eta}{3} T}{\eta T} \right)^3 - 15 \left(\frac{t - \frac{1-\eta}{3} T}{\eta T} \right)^4 + 6 \left(\frac{t- \frac{1-\eta}{3} T}{\eta T} \right)^5 + \frac{60 \left( \left(\frac{t-\frac{1-\eta}{3} T}{\eta T} \right) - 3 \left(\frac{t-\frac{1-\eta}{3} T}{\eta T} \right)^2 + 2 \left(\frac{t-\frac{1-\eta}{3} T}{\eta T} \right)^3 \right)}{(\eta T)^2 \left( \frac{A_{st}}{m \sigma_{st}^2} + \frac{A_{mt}^{max,cr}}{m \sigma_{mt}^2} \right)} + \frac{4 A_{st}}{d^2 \left( \frac{A_{st}}{\sigma_{st}^2} + \frac{A_{mt}^{max,cr}}{\sigma_{mt}^2} \right)} }{1+\frac{8 A_{st}}{d^2 \left( \frac{A_{st}}{\sigma_{st}^2} + \frac{A_{mt}^{max,cr}}{\sigma_{mt}^2} \right)}} \, .
\end{equation}
\end{widetext}
A direct evaluation of the coefficients of the second and third term of the numerator as well as the second term in the denominator leads to values below $10^{-3}$, meaning that the STA trajectory closely resembles the minimum jerk one. Importantly, with the same Taylor expansion we also derived an approximation for the maximum amplitude of the STA pulse in terms of the maximum amplitude over the capturing or releasing stage (or, equivalently, the initial amplitude in the transport interval), $A_{mt, \text{STA}}^{max} \approx A_{mt}^{max,cr}+ A_{st} (1+2/e^{3/2})  \sigma_{mt}^2/\sigma_{st}^2 \geq A_{mt}^{max,cr}$. The latter expression allows us to compare pulses with the same global maximum amplitude, which is the relevant experimental parameter. The approximations obtained for the STA control pulses using the same experimentally realistic values for the parameters as in the previous sections are depicted in panels (c) and (d) of Fig. \ref{fig:ts_02} as dashed orange lines. We checked that the approximation for the maximum amplitude of the STA ramp presents a very good agreement with the full numerical results. We also corroborated that the relative error between the full numerical solution and the approximation for the amplitude is less than $4\%$ in all the duration of the transport interval. 

\begin{figure*}[htb]
    \centering
    \includegraphics[width=\textwidth]{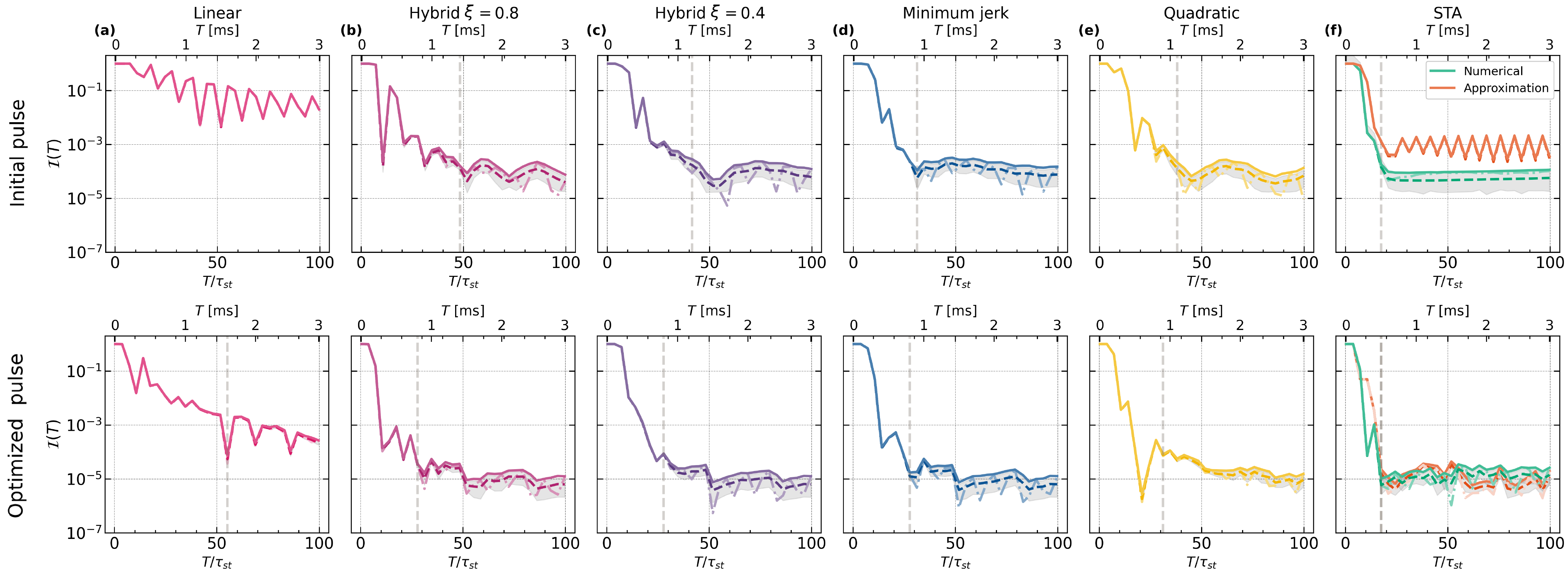}
    \caption{Error after transport. Infidelity $\mathcal{I}$ versus total time $T$ for the piece-wise linear (a), hybrid with $\upxi=0.8$ (b) and $\upxi=0.4$ (c), minimum jerk (d), quadratic (e), and STA (f), for the initial pulses (top) and optimized ones (bottom). All the pulses have a maximum amplitude of $A_{mt}^{max}/\hbar= 3.57 \times 2\pi \text{ MHz}$, the transport time is given by $\eta \, T$ with $\eta=2/5$, and the total covered distance between static tweezers is $d = 7\, \mu\text{m}$. The maximum, average, and last value of the infidelity over the last stage of the protocol [see Eqs.~\eqref{A_mt_of_t} and \eqref{x_mt_of_t}] are shown as solid, dashed, and dotted lines respectively. The dashed area indicates the standard deviation calculated over the same time interval and quantifies the error in the infidelity. The vertical gray dashed line highlights the value of the total time for which the infidelity reaches the threshold of $10^{-4}$. For the hybrid pulse with $\upxi=0.8,\, 0.4$, minimum jerk, quadratic and STA ramps those times are respectively of about $ 48,\, 41,\, 31,\, 38,\, 18 \,\tau_{st}$, with $\tau_{st}$ being the characteristic time of the static traps (the characteristic time during transport is $\tau_{mt} \approx \tau_{st}/2$). For the two hybrid and minimum jerk optimized pulses these times decrease to $28 \, \tau_{st}$, the time for the quadratic ramp decreases to $31 \,\tau_{st}$, the STA threshold time does not change, and the piece-wise linear optimized pulse reaches the infidelity threshold for a total time of about $55 \,\tau_{st}$.}
    \label{fig:ts_03}
\end{figure*}

In a recent manuscript, Hwang \emph{et al.} experimentally demonstrate the advantage of an STA-based trajectory over the constant velocity and constant jerk trajectory \cite{hwang_arxiv_2024}. By using also STA techniques and a harmonic truncation of the tweezer potential, the team of Jaewook Ahn developed a very similar trajectory to the one given in Eqs.~\eqref{eq_x_mt_approx} and \eqref{eq_x_mt_approx_2}, and beautifully prove that the STA survival probability of atoms after transportation outperforms the non-STA ones. We would like to highlight that even though our approaches are similar there are two key differences; first, their calculations consider only the moving tweezer and not the static ones, and second, they do not develop the amplitude pulse to make use of the depth of the tweezer as a second control variable. As we show in the next section, the incorporation of a modulation in the amplitude of the moving tweezer that counteracts the restoring force of the static traps allows the STA pulse to outperform the experimentally motivated ones even without optimization. We trust that the simple approximations for our STA solutions provided in this section will motivate their experimental implementation in the near future. 

\section{Atom transport characterization and optimization}
\label{sec_transport_char_opt}

In this section we present the performance analysis of the considered pulses and their subsequent optimization.

\subsection{Performance of the initial guess pulses}
\label{sec_perf_no_opt}

In order to evaluate the performance of the transport of a single atom under the piece-wise linear, hybrid, minimum jerk, quadratic, and STA pulses, we are mainly concerned with two features; first, that the transport is faithful (fidelity condition), and second, that the vibrational excitations of higher states are reasonably bounded to prevent atom losses \cite{savard_pra_1997, gardiner_pra_2000}. To check the first condition we analyze the transport infidelity.  As explained in Sec.~\ref{sec_model}, the infidelity quantifies the error after the transport and it is closer to zero when the state of the atom is closer to the target state. Since our target state is the ground state of the target tweezer, a lower infidelity at the end of the transport is equivalent to a no-heating condition after the complete transport protocol.

Since the potential of an optical tweezer has finite depth, it is also necessary to assess whether the no-heating condition is also satisfied during transport. As discussed in Ref.~\cite{gardiner_pra_2000}, the main effect of heating is to expel the atom from the trap not as a result of an increase in the mean energy but as a consequence of the spreading of the width of the distribution over the energy states (the physical picture is that when the upper tail of the distribution reaches untrapped levels the atom is lost). Based on this argument, in order to quantify the increase in the vibrational quantum number we calculate the mean value $\expval{N}$ and the width $\Delta N$ of the distribution of the atomic state over the instantaneous eigenstates of the moving tweezer. Following Ref.~\cite{hickman_pra_2020}, we also calculate an effective temperature given by the expectation value of the kinetic operator $T_\text{eff}=2\expval{K}/k_B$, where $k_B$ is the Boltzmann constant and $K = p^2/(2m)$. By means of the virial theorem for a harmonic trap we have $\expval{H} = 2 \expval{K}$ \cite{Cohen-Tannoudji}, therefore $T_\text{eff}$ is a measure of the mean energy of the system which is usually reported to analyze the stability of the system in transient times \cite{zhang_pra_2015, guery_odelin_pra_2014}. In our case, the maximum depth of the moving tweezer during transport hosts about 55 oscillator levels and the deviation of the lower states with respect to the harmonic ones is small. In what follows we use SI and characteristic units for the time; from the frequency of the static potential [see Eq.~\eqref{eq_omegast}] we can define a characteristic time as $\tau_{st}=2\pi/\omega_{st} \approx 0.03 \text{ ms}$. We use this value since it is fixed for any maximum amplitude of the moving tweezer, however, the relationship between this value and the characteristic time during transport is $\tau_{mt} = 2\pi/\omega_{mt}^{max} \approx \tau_{st}/2$.  

The infidelity as a function of the total time $T$ is reported in Figure \ref{fig:ts_03} for the piece-wise linear, hybrid ($\upxi =0.8,\, 0.4$), minimum jerk, quadratic, and STA (full numerical solution and approximation) pulses from left to right. We consider total times ranging between $0.01$ and $3 \text{ ms}$ in agreement with the experimental values (see for instance Ref.~\cite{liu_prx_2019} or \cite{spence_njp_2022}). We calculate the maximum (solid curve), average (dashed line) and last value (dotted line) of the infidelity over the final interval of the pulse. Since it constitutes an upper bound, we propose the maximum value as the quantity to be used when comparing with experimental results. The shaded area indicates one standard deviation as a measure of the error in the fidelity. 

As expected, the piece-wise linear pulse produces the highest error after transport (infidelity higher than $10^{-2}$ for all $T$) and sustained oscillations (with a period of about $12 \, \tau_{st}$). The hybrid pulse behavior interpolates between the minimum jerk and the linear one and presents more oscillations for higher $\upxi$ (i.e.\ going towards the linear ramp, as expected). The quadratic pulse infidelity has a similar behavior to the one of the hybrid pulse with $\upxi \lesssim 0.5$ but with a slightly higher oscillatory behavior. Among all the pulses, the minimum jerk and the STA ones present the smoothest and most stable behavior. Moreover, the STA pulse also allows to reach the smallest infidelity in the fastest time; the infidelity rapidly decreases for $T$ between $10$ and $20 \, \tau_{st}$ and then reaches a stable value around $10^{-4}$. In general, we observe that all the pulses fail in the transport task for $T \lesssim 10 \, \tau_{st}$. For total times larger than $10 \, \tau_{st}$ the infidelity decreases when the total time increases and the hybrid, minimum jerk, quadratic, and STA pulses present two more regimes: a diabatic regime with some oscillations between $T \approx 10 \, \tau_{st}$ and a value in the range between $20$ to $50 \, \tau_{st}$, and the adiabatic regime for sufficiently large $T$. The total time where the second regime ends depends on the pulse and corresponds in increasing order to the STA, minimum jerk, hybrid with $\upxi=0.4$, quadratic, and hybrid with $\upxi =0.8$ ramps. After that, the fidelity saturates with some smooth oscillations that improve when the upper time threshold of the diabatic regime decreases. Interestingly, the STA approximation captures well the two first regimes, including the value of the diabatic threshold but presents strong oscillations with a period of about $8 \, \tau_{st}$ in the last adiabatic regime. For total times $T \gtrsim 50 \, \tau_{st}$, the hybrid, minimum jerk, quadratic, and STA pulses reach the adiabatic regime and therefore any of those pulses can faithfully perform the transport task. Since for short $T$ the STA ramp presents a clear advantage, we conclude that our proposed STA protocol has the best performance, achieving the infidelity threshold of $10^{-4}$ (vertical dashed line) in nearly half of the time taken by the minimum jerk trajectory (second best). In Appendix \ref{sec_app_transport_evol} we present examples of the evolution of the infidelity for some particular values of total pulse durations $T= 0.5, \, 1.5,\, 2.5,\, 3 \text{ ms}$. 

In order to test the performance of the STA solutions in a different experimental setting, we compute the pulses and simulate the evolution using the parameter values described in \cite{schymik_PRA_2020}. Both the numerical and approximate STA protocols we propose complete the transport and transfer between two neighbouring tweezers separated by a distance of $5 \text{ $\mu$m}$ in $160 \text{ $\mu$s}$ with an infidelity of about $10^{-4}$. Comparing this time to a total time of $1.25 \text{ ms}$ reported in Ref.~\cite{schymik_PRA_2020} our pulse performs the task $7.8$ times faster even when the transport stage is done slower (for us it takes $80 \, \mu\text{s}$ while in the experiments of Ref.~\cite{schymik_PRA_2020} it takes $50\,\mu\text{s}$). More strikingly, our STA protocols reduce by a factor of 15 the time devoted to capturing or releasing (in our case $40 \, \mu\text{s}$ versus $600\,\mu\text{s}$ for the experiments reported in Ref.~\cite{schymik_PRA_2020}). We would like to stress that these results could be further improved by adapting the ratio between the time devoted to the transport and transfer stage to parameters of the system at hand.

\begin{figure}[htb]
    \centering
    \includegraphics[width=0.45\textwidth]{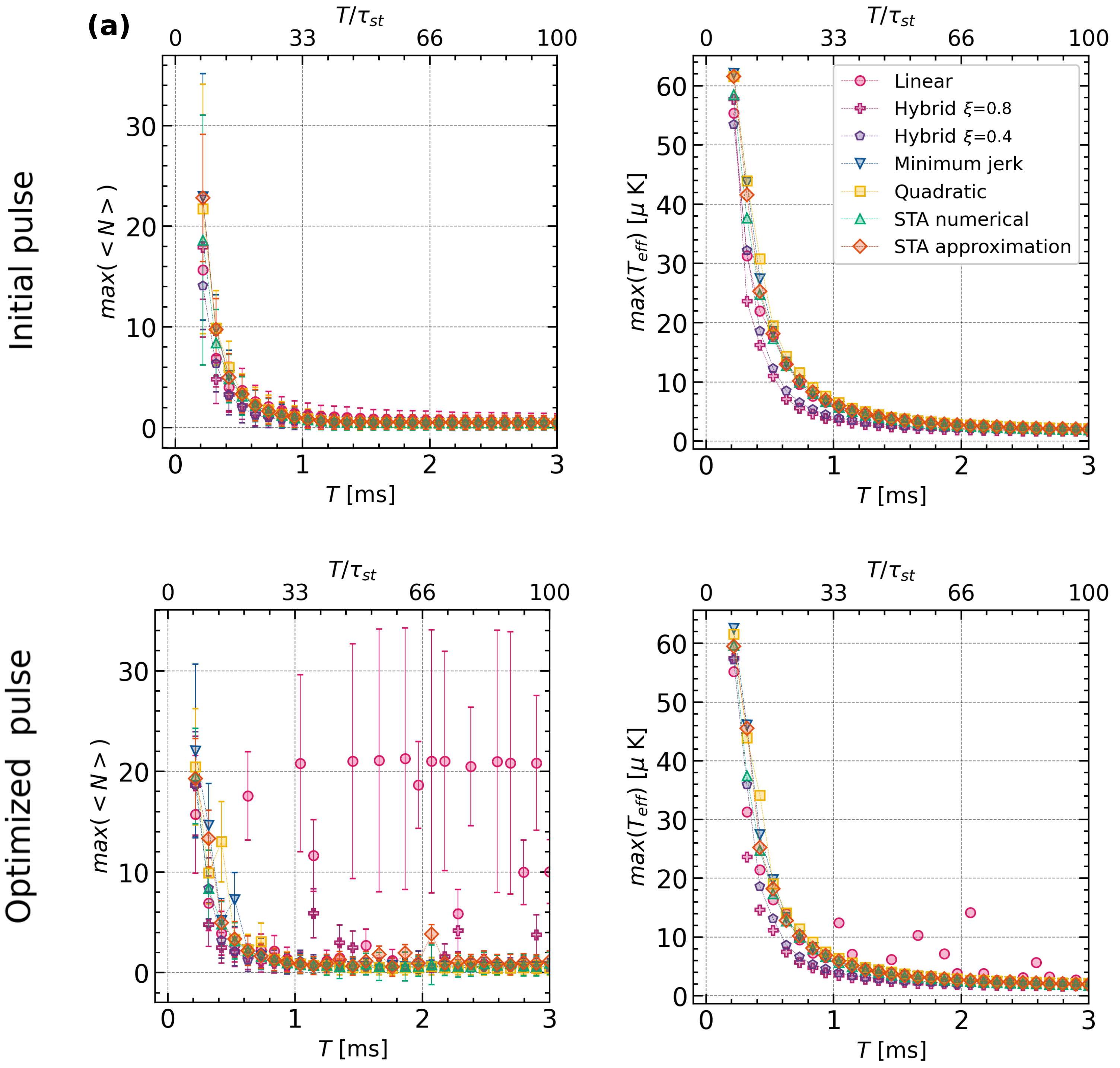}
    \caption{Upper bound measures of transient heating. (a) Maximum mean occupied level (points) and width (error bars) of the distribution over the moving tweezer states (denoted respectively by $\expval{N}$ and $\Delta N$) and (b) maximum effective temperature $T_\text{eff}$ over the complete duration of the pulse as a function of the total time of the pulse $T$ for the initial pulses (top row) and the optimized ones (bottom row). The measures obtained for the piece-wise linear, hybrid with $\upxi = 0.8, \, 0.4 $, minimum jerk, quadratic, and STA (fully numerical solution and approximation) ramps are depicted as magenta circles, wine crosses, purple pentagons, blue down triangles, yellow squares, turquoise up triangles, and orange diamonds, respectively,  with a line of the same color as a guide for the eye. For a total pulse time of $T \approx 0.22 \text{ ms} \approx 8 \, \tau_{st}$ the mean occupied level plus its uncertainty equals half of the states hosted by the moving tweezer. All the initial pulses have a maximum amplitude of $A_{mt}^{max}/\hbar=3.57 \times 2\pi \text{ MHz}$, a transport time of $\eta \, T$ with $\eta=2/5$, and a total covered distance between static tweezers of $d = 7\, \mu\text{m}$. As before $\tau_{st}$ denotes the characteristic time of the static traps, related to the characteristic time during transport via $\tau_{mt} \approx \tau_{st}/2$.}
    \label{fig_dN_Teff}
\end{figure}

Figure \ref{fig_dN_Teff} depicts the maximum expected value $\text{Max(}\expval{N}\text{)}$ and the associated uncertainty or width of the distribution $\text{Max(}\Delta N\text{)}$ for the occupied states of the moving tweezer together with the the maximum effective temperature $\text{Max(}T_\text{eff}\text{)}$ as a function of the total pulse time $T$ \footnote{The expected value and maximum width of the distribution over the moving tweezer states are computed on the transport interval as follows. For each time $t$ we compute the states $\ket{\phi_n(t)}$ of the moving trap via diagonalization of the moving tweezer Hamiltonian, then we project the evolved wave-function $\ket{\psi(t)}$ over each of these states to calculate the weights or populations over each state $p_n = |\braket{\phi_n(t)}{\psi(t)}|^2$. Once we have the population distribution $\{p_n\}$ we calculate the variance and mean value as usual. We performed the same calculation by using a harmonic approximation of the moving trap and we check the consistency of the results. On the other hand, the maximum effective temperature $\text{Max(}T_\text{eff}\text{)}$ is calculated over the complete protocol duration.}. As expected, both quantities increase for shorter $T$ because diabatic processes induce a higher mixture of states. The increase in the vibrational modes shows a similar behavior for all the considered pulses; for total times above $0.5 \text{ ms}$ the wave function has a dominant weight over the ground state of the moving trap and smaller weights over the first and second excited states. For total pulse times shorter than $0.5 \text{ ms}$ the expectation value of the occupied moving tweezer states $\expval{N}$ and its uncertainty $\Delta N$ increase rapidly for decreasing total times. For $T \approx 0.22 \text{ ms} \approx 8\, \tau_{st}$ the mean occupied level plus the associated uncertainty reaches half of the trap states, i.e. 27 levels for our trap hosting a total of 55 states. An estimation of the vibrational number increase for a moving harmonic oscillator can be obtained by evaluating the Fourier transform of the acceleration at the frequency of the oscillator \cite{guery_odelin_pra_2014, carruthers_ajp_1965, bluvstein2022quantum}. By computing the Fourier transform of the acceleration of our approximated STA trajectory [Eq.~\eqref{eq_x_mt_approx_2}] we find an agreement with less than 10\% of relative difference between the time at which our protocol reaches an increase of 27 oscillator levels and the time that would be needed by the STA solution proposed by the group of Jaewook Ahn, see Eq. (7) of Ref.~\cite{hwang_arxiv_2024}.     

The hybrid ramps demonstrate a slight advantage when compared to the rest of the pulses, as for total times $T \lesssim 1.5 \text{ ms}$ the hybrid pulses present a smaller effective temperature. Also, the difference in the effective temperature among the hybrid ramps and the other pulses increases for shorter total times reaching a difference of about $22 \, \mu\text{K}$ for $T \approx 0.3 \text{ ms}$ between the quadratic and the hybrid ramp with $\upxi = 0.8$. Since the change of the energy after the evolution of a moving harmonic oscillator for a particle initially in one of the oscillator states is given by the the Fourier transform of the acceleration evaluated at the frequency of the oscillator \cite{guery_odelin_pra_2014, carruthers_ajp_1965, bluvstein2022quantum}, a purely linear pulse does not induce heating during transport. The incorporation of the linear part in the hybrid trajectory presented in Ref.~\cite{liu_prx_2019} uses the latter advantage and at the same time reduces the error after transport (as discussed on the basis of Fig. \ref{fig:ts_03}, the piece-wise linear pulse infidelity is more than two orders of magnitude higher when compared to the other pulses) induced by the velocity discontinuities at the endpoints of the pulse by replacing them with the minimum jerk sub-intervals. In Appendix \ref{sec_app_transport_evol}, we show $\expval{N}$, $\Delta N$, and $T_\text{eff}$ as a function of time for total protocol times $T= 0.5, \, 1.5,\, 2.5,\, 3 \text{ ms}$, where it is possible to see that for $T = 0.5 \text{ ms}$ the hybrid pulse with $\upxi = 0.8$ has the smaller $\Delta N$ and $T_\text{eff}$, followed by the hybrid pulse with $\upxi =0.4$. 

\subsection{Atom transport optimization}
\label{sec_transport_opt}

To further reduce the transport error we implement an optimization protocol using the d-CRAB algorithm introduced by Rach \emph{et al.} in 2015 \cite{caneva2011chopped, doria2011optimal, rach2015dressing, muller2022one}. This algorithm offers several advantages over other optimization methods, mainly because of its efficiency managing high-dimensional control spaces; by expanding the control pulses in a randomized basis, it reduces the dimensionality of the search space while circumventing local traps that could arise due to this restriction. Since within d-CRAB only a small number of parameters are optimized at the same time, the need for gradient information is mitigated \cite{bergholm2019optimal, khaneja2005optimal, machnes2011comparing}. This makes it possible to perform optimal pulse shaping adaptively by using experimental data, which allows to implicitly take into account experimental unknowns such as parameter uncertainties and instrumentation transfer functions \cite{Singh2023} in the optimization cycle. We observed that in practice the flexibility offered by d-CRAB comes at the price of an increased dependence on the initial guess for the controls, likely the result of a more local search in control space. In light of this, the in-depth discussion of the relevant experimentally feasible pulses and STA protocol that we presented in Sec.~\ref{sec_pulses} becomes especially relevant. 

We use the QuOCS library \cite{rossignolo2023quocs}, which incorporates, among others, the d-CRAB algorithm in a user-friendly interface \cite{rach2015dressing}. The QuOCS library allows for a straightforward setting of the parameters related to the pulse, such as total time and basis parameters. It also allows to implement scaling functions that set appropriate limits on the position and amplitude control. We use an expansion of the control pulse in a basis of Sigmoid functions and select only some of the coefficients as optimization control. We then conduct a systematic optimization involving four different optimization combinations. Following the methodology outlined in \cite{Lam2021}, we optimize the trajectory for different total times and adjust the initial controls to identify the shortest achievable time during the numerical analysis. 

The infidelity after transport as a function of the total time $T$ for all the considered pulses is shown in Fig. \ref{fig:ts_03} (bottom row). The optimization improves the fidelity by two orders of magnitude for the linear pulse and by one order of magnitude for the remaining pulses. For all the pulses, the standard deviation over the last interval of the pulse (shaded area) is reduced by the optimization. The distance between the maximum (solid line) and average value (dashed) over the last stage of the pulse is also reduced. All of this points towards a much more stable state after transport, see also Fig. \ref{fig_inf_evol} of Appendix \ref{sec_app_transport_evol}. The piece-wise linear optimized pulse reaches the infidelity value of $10^{-4}$ for a total time pulse of $55 \, \tau_{st}$. This time remains at $18 \, \tau_{st}$ for the STA pulse, not changing appreciably after the optimization. The threshold time for the remaining pulses decreases between 10 and 30\%; for the hybrid pulse with $\upxi = 0.8$ it changes from $48$ to $28 \, \tau_{st}$, for the hybrid ramp with $\upxi =0.4$ from $41$ to $28 \, \tau_{st}$, from $32$ to $28\, \tau_{st}$ for the minimum jerk ramp, and for the quadratic pulse the threshold time is reduced from $38$ to $31\, \tau_{st}$ (all these times are indicated with a dashed gray vertical line). We conclude that the optimization yields a significant improvement for all the pulses. However, we would like to highlight that the full numerical solution for the STA protocol shows highly desirable features even without optimization. In particular, our non-optimized STA solution implies an improvement of 42\% in the total time required to reach the fixed infidelity threshold of $10^{-4}$ with respect to the non-optimized minimum jerk pulse. In the case of the optimized ramps, the STA solution improves that time by 36\% with respect to the optimized minimum jerk pulse.

As we mentioned before, since the objective of the optimization is to minimize the final infidelity, there is no guarantee that during the obtained transport dynamics high-energy instantaneous eigenstates are not populated, allowing the particle to escape the trap. For this reason, it is very important to check that the optimized pulses satisfy the no-heating condition for transient times. As can be seen in Fig. \ref{fig_dN_Teff} (bottom row), the optimization leads to higher vibrational excitations for the piece-wise linear and hybrid pulse with higher hybridization ($\upxi=0.8$). This feature is revealed when inspecting the effective temperature but it is magnified by the behavior of the mean occupied level and the width of the distribution over the oscillator states, in line with the discussion presented in Ref.~\cite{gardiner_pra_2000}. Our results suggest to avoid the use of piece-wise linear pulses which are known to induce heating due to their intrinsic velocity discontinuities, this is consistent with the good fidelity obtained in Ref.~\cite{spence_njp_2022} by using a pulse with constant velocity in the central 10\% of the transport interval ($\upxi = 0.1$). The evolution of the infidelity, $\expval{N}$, $\Delta N$ and $T_\text{eff}$ is shown in Appendix \ref{sec_app_transport_evol} for total times $T= 0.5, \, 1.5,\, 2.5,\, 3 \text{ ms}$. The appendix also contains a comparison between some examples of initial and optimized pulses. We would like to mention that a very small change in the pulse can translate into a considerable improvement of the fidelity, as was already exposed from the difference in the performance of the full numerical solution for the STA pulse when compared to the behavior of the approximated solution. Since all the optimized pulses except the linear one satisfy the no-heating condition, our results suggests that the usage of d-CRAB for pulse shaping would not result in collateral heating that would cause major atom loss during transport.

\begin{figure}
    \centering
    \includegraphics[width=0.50\textwidth]{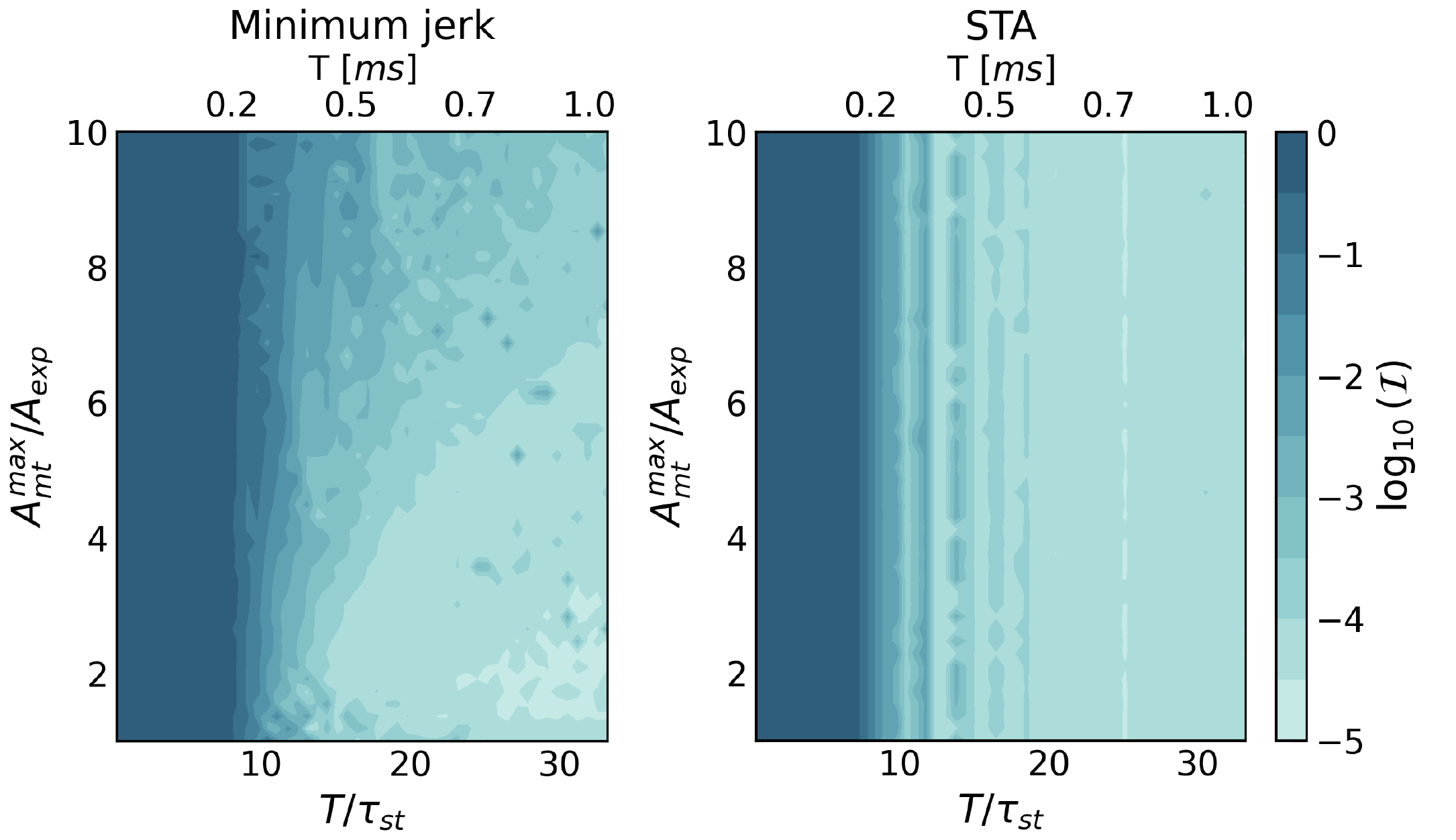}
    \caption{Determining optimal regions in the experimentally accessible parameters space for high fidelity transport pulses. Infidelity obtained for the optimized minimum jerk ramp (left) and for our proposed STA pulse also after optimization (right) as a function of the maximum amplitude $A_{mt}^{max}$. The amplitude is in units of $ A_{exp}/\hbar = 3.57 \times 2\pi \text{ MHz}$ and the total time of the pulse $T$ is in units of the characteristic time of the static traps $\tau_{st}$ (bottom scale) and in ms (upper scale). Lighter colors show regions of smaller infidelities which quantifies the error in the state obtained after transport. As before, the characteristic time during transport is given by $\tau_{mt} \approx \tau_{st}/2$.}
    \label{fig_phase_diag}
\end{figure}

As a final goal we turn into the determination of good regions in the parameter space where the transport protocol can be reliably implemented in current experiments. We therefore focus on the two parameters that can be changed easily in the experiments, namely the total time of the pulse and the maximum amplitude (depth) of the moving tweezer. In fact, while the amplitude can be experimentally adjusted by manipulating the laser power, other parameters (such as the width of the tweezer) might be challenging to modify since they are ultimately determined by the optical elements within the apparatus. In order to identify good intervals for the amplitude and total time to run the experiments, we report in Fig. \ref{fig_phase_diag} the heat maps for the infidelity (maximum over the last interval of the pulse) for the two pulses that enable to reach the infidelity threshold of $10^{-4}$ in the shortest total time for a fixed amplitude of $A_{mt}^{max}/\hbar = A_{exp}/\hbar = 3.57 \times 2\pi \text{ MHz}$, i.e.\ the optimized minimum jerk and STA pulses. We consider total times up to $1 \text{ ms} \sim 33 \, \tau_{st}$ (for larger total times the infidelity reaches the stable regime) and maximum amplitudes between $A_{exp}$ and $10 \, A_{exp}$, both being realistic ranges of the parameters for experimental realizations. Our simulations show that the minimum jerk pulse infidelity has a strong dependence on both the total time and the maximum amplitude. For larger amplitude values, larger times are needed in order to achieve a desired fixed value of the infidelity. We interpret this on the basis of the relation between the capturing/releasing time and the total time of the transport, here fixed to $T/5$ and $2\,T/5$ respectively. A higher amplitude of the moving tweezer implies a larger difference between the moving and static potential meaning that the system requires a larger releasing time in order to adjust towards the target state. Fixing an error threshold of $10^{-4}$, the best region on the parameter space appears for times longer than $20 \,\tau_{st}$ and amplitudes between $1$ and $4 \, A_{exp}$. On the other hand, by incorporating a modulation in the amplitude of the moving tweezer that counteracts the restoring force of the static traps, the STA pulse depicts a quite robust infidelity against amplitude variations. Using our STA pulse, infidelities below $10^{-4}$ can be obtained for times longer than $20 \,\tau_{st}$ independently of $A_{mt}^{max}$. The STA ramps also seem to present ``magic'' times windows exposed as lighter vertical stripes indicating that the infidelity is suppressed by at least one order of magnitude around $T \approx 10.5,\, 13.5,\, 15.5,\,17.5,\, 25 \, \tau_{st}$ \cite{Zhang2016}. 

Figure \ref{fig_phase_diag} shows a drastic increase in the infidelity for total pulse durations below $8 \, \tau_{st}$. This feature can be already inferred from Fig. \ref{fig:ts_03} for all the considered pulses and for $A_{mt}^{max} = A_{exp}$. Our results strongly point towards the presence of a quantum speed limit (QSL), imposing a bound for the minimum time required for the transport task \cite{caneva2009optimal, deffner_jpa_2017}. Taking into account that the transport takes $2/5$ of the total time, the bound for the transport time is about $3\, \tau_{st}$. When comparing this value for the same distance in oscillator units and for a trap hosting the same amount of states, we observe that our lower bound for the transport time is five times larger than the one reported in Ref.~\cite{pagano_prr_2024} for atom transport in an optical tweezer setup with a quadratic pulse. However, it is important to highlight that our protocol incorporates more information processing details than the one considered in Ref.~\cite{pagano_prr_2024}, where only the transport stage is taken into account with no consideration of the time devoted to capture and release the atom. For a moving trap hosting the same amount of levels, we obtain a lower time threshold for the transfer between tweezers which is 9 times faster than the time reported in Ref.~\cite{spence_njp_2022} (8 oscillator units versus 75 oscillator units). Considering that in Ref.~\cite{schymik_PRA_2020} the authors state that the time needed for the capturing or releasing is 12 times the one required for the transport task, our results suggests that an improvement of the transfer between tweezers can translate into an overall speedup even at the cost of devoting a larger time to the transport stage. Our findings constitute a wake-up call towards the importance of including the transfer between tweezers in the model; by optimizing over the entire process, which also includes the effect of the static traps, we see the importance of tailored solutions, such as those based on our specialized STA, for reaching error rates in the fault tolerant regime.

\section{Summary and Conclusions}
\label{sec_concl}

In this work we focused on the transport and transfer between tweezers of neutral atoms, a relevant problem (and possibly a bottleneck) for the creation and improvement of quantum processors and simulators \cite{bluvstein2022quantum,bluvstein2024logical}. We considered four different kinds of experimentally motivated pulses: piece-wise linear, piece-wise quadratic, minimum jerk, and a family of hybrid linear and minimum jerk ramps. These pulses were used to transport not only atomic ensembles but also few and single atoms in experiments with optical tweezers \cite{schymik_PRA_2020, Endres2016, bluvstein2022quantum, bluvstein2024logical, liu_prx_2019, picard_prx_2024, zhang_qstiop_2022} and optical conveyor-belts \cite{matthies_thesis_2023}. To generate a pulse that is specifically tailored for our problem, we developed a transport protocol using Shortcuts to Adiabaticity (STA) methods. The main difference between our STA approach and previous approaches is that we take into account the potential of the static tweezers to optimize the transfer between static and moving traps. In order to facilitate the adoption of our pulse in future experiments, we provided a handy and easy to implement approximation for the full numerical solution. A close inspection of the approximated STA solution shows that the associated trajectory is a modification of the widely known minimum jerk trajectory with the key difference that its amplitude modulation (related to the depth of the moving trap) counteracts the effect of the static tweezers. We characterized the performance of all the considered pulses and found that our proposed STA protocol outperforms the experimentally inspired ramps even without optimization. 

All the considered pulses were used as initial guesses for QuOCS optimization, leading to a reduction in the error (quantified by the infidelity with respect to the target state) by at least one order of magnitude. By setting a threshold for the infidelity of $10^{-4}$ we obtained a compression of the total time of the pulse between 10\% and 30\% after optimization for the hybrid, minimum jerk, and quadratic pulses. 

Based on the discussions on heating and losses presented in Refs.~\cite{bluvstein2022quantum, hickman_pra_2020}, in order to check that our optimized pulses do not induce extra heating during transport, we calculated the mean occupied level and the width of the distribution over the states of the moving tweezer together with an effective temperature related to the motional energy. For total pulse times below $8 \, \tau_{st} \approx 16 \,\tau_{mt}$ (with $\tau_{st}$ and $\tau_{mt}$ being the characteristic time of the static and moving traps respectively), the increase in the vibrational excitations exceeds half of the states hosted by the moving trap. We focused on the two pulses with the best performances, namely the minimum jerk and STA, and by quantifying the error after transport as a function of the total time of the pulse and the maximum depth of the moving tweezer (both experimentally accessible parameters) we identified good regions in the parameter space to reliably run the transport protocol. For total times larger than $10\,\tau_{st}$ both pulses provide a good performance (infidelity below $10^{-2}$), however, the minimum jerk ramp entails limitations on the maximum amplitude that can be selected while our proposed STA pulse provides much more freedom. We conclude that our STA solution constitutes an improvement compared to the usual minimum jerk trajectory. 

Our findings also suggest the presence of a lower bound for the total time of the pulses; we obtain a numerical quantum speed limit for the complete protocol time of about $8 \, \tau_{st}$. Since the transport time is a fraction of the total duration of the protocol, this value corresponds to a bound of $3\, \tau_{st}$ for the transport stage, which is around five times the one reported in Ref.~\cite{pagano_prr_2024} for a protocol that does not considers the time required to capture and release the atom. Taking into account that the capturing or releasing time was estimated to be 12 times larger than the time needed for the transport process only in the experiments of Ref.~\cite{schymik_PRA_2020}, and that our obtained transfer time is 9 times faster than the one reported in Ref.~\cite{spence_njp_2022}, we interpret our results as a call to attention towards the relevance of including the transfer between tweezers in the model. In other words, to fully capture the information protocol it is necessary to consider not only the transport process but also the transfer between optical tweezers, which makes up a significant portion (if not the majority) of the error and the time budget. 

Our analytical and numerical results show that small deformations in the pulses (for instance, less than 4\% for the approximated STA amplitude when compared to the full numerical solution) can translate into a decrease in the error after transport by two orders of magnitude. We also contribute to fill a knowledge gap identified in Ref.~\cite{matthies_thesis_2023} in the case of atomic transport by means of optical conveyor-belts, namely providing a systematic characterization of the effectiveness of different trajectories. 

Our proposed STA pulse also proves that a modulation of the depth of the tweezer different from the widely used linear ramps and, especially, the addition of a term that counteracts the effect of the static tweezers, produces smaller and more stable errors for shorter pulse durations. It is worth noticing that to the best of our knowledge our proposed amplitude could be implemented in current experiments since, as mentioned in Ref.~\cite{bluvstein2024logical}, amplitudes with a quadratic dependence on time are possible. The implementation of the modulation in the tweezer depth as a second control is particularly promising in light of the recent experimental implementation of an STA trajectory which has many common elements to ours \cite{hwang_arxiv_2024}, with the key difference given by correction of the effect of the static tweezers and the transfer process. Our results contribute to the generalization and improvement of previously known ramps for transporting neutral atoms in state-of-the-art quantum processors and quantum simulators based on tweezers arrays. Given the flexibility of our approach to accommodate for different external potentials, we are currently working on its application to optical lattice platforms.

\begin{acknowledgments}
We thank Thomas Reisser for technical support related to QuOCS. We thank Jan Reuter and Matteo Rizzi for discussions on the numerical stability.
We also thank Christian Gro\ss, Philip Osterholz, Yu Hyun Lee, Peter Bojovic, and Titus Franz for valuable discussions on the experimental possibilities. 
We acknowledge funding from the German Federal Ministry of Education and Research through the funding program quantum technologies—from basic research to market under the project FermiQP, \href{https://www.quantentechnologien.de/forschung/foerderung/quantenprozessoren-und-technologien-fuer-quantencomputer/fermiqp.html}{13N15891} and under Horizon Europe programme HORIZON-CL4-2022-QUANTUM-02-SGA via the project 
\href{https://doi.org/10.3030/101113690}{101113690} (PASQuanS2.1). 
E.C. was supported by JSPS KAKENHI grant number JP23K13035. 
\end{acknowledgments}


\appendix

\section{STA Theory}
\label{sec_app_sta}
\renewcommand{\theequation}{A.\arabic{equation}}
\renewcommand{\thefigure}{A.\arabic{figure}}
\setcounter{figure}{0}
We regroup and summarize here for the convenience of the reader the well known results from the theory of Shortcuts to Adiabaticity (STA) which are relevant in the derivation of the STA-based transport pulse in Sec.~\ref{sec_pulses_sta}.

\textit{Dynamical invariants and STA.} Shortcuts to Adiabaticity are a collection of methods whose aim is to obtain a fast-forward version of the adiabatic time evolution of the system. The strategy we employ is to exploit a known dynamical invariant of the system to reverse engineer a STA pulse that realizes the desired dynamics. Let us start by considering a time dependent Hamiltonian $H(t)$ and the solutions $\ket{\psi(t)}$ to the Schr{\"o}dinger equation
\begin{equation}
i\hbar\frac{\partial }{\partial t}\ket{\psi(t)} = H(t)\ket{\psi(t)}.
\label{eq:schr}
\end{equation}
A dynamical invariant of $H(t)$ is a time-dependent operator $I(t)$ that satisfies the relation
\begin{equation}
\frac{d I}{dt} \equiv \frac{\partial I}{\partial t} + \frac{1}{i\hbar}\left[I,H\right] = 0,
\label{eq:inv}
\end{equation}
meaning that $\frac{d}{dt} \ev{I(t)}{\psi(t)}$. As long as $I^{\dagger}=I$ is hermitian, it is possible to write any solution of Eq.~\eqref{eq:schr} as a time-independent linear combination of eigenvectors $\ket{\phi^{LR}_n(t)}$ of $I(t)$ \cite{LR1969}, which we call dynamical modes 
\begin{align}
    & \ket{\psi(t)} = \sum_n c_n \ket{\phi^{LR}_n(t)},  \text{ with } \nonumber \\
    & I(t)\ket{\phi^{LR}_n(t)} = \lambda_n \ket{\phi^{LR}_n(t)} \,,
\end{align}
and where we explicitly state that the eigenvalues $\lambda_n$ of $I(t)$ do not depend on time. Additionally, in the case that $I(t)$ does not involve time-differentiation, the dynamical modes can be explicitly constructed by means of a gauge transformation of the (generic) eigenvectors $\ket{\phi_n(t)}$ of $I(t)$ \cite{LR1969}, featuring the so-called Lewis-Riesenfeld phases $\gamma_n(t)$:
\begin{align}
\ket{\phi^{LR}_n(t)} &= e^{i\gamma_n(t)}\ket{\phi_n(t)},\\
\hbar \frac{d\gamma_n}{dt} &= \bra{\phi_n(t)}i\hbar\frac{\partial }{\partial t} - H(t)\ket{\phi_n(t)}.
\end{align}

In addition to being a useful tool to solve the equation of motion Eq.~\eqref{eq:schr}, these modes can be used to find the Hamiltonian operator that produces a certain prescribed time-evolution by means of reverse engineering. The goal is to realize a time evolution that maps eigenstates of $H(t_i)$ to eigenstates of $H(t_f)$ in a given amount of time $ T = t_f-t_i$ (the subindices $i,f$ denote, as usual, initial and final). One way to obtain this is to impose that every dynamical mode coincides (up to a phase factor) with an eigenstate of the Hamiltonian at $t=t_i,t_f$. This requires the dynamical invariant to commute with the Hamiltonian at the endpoints of the time evolution, leading to the following boundary conditions 
\begin{equation}
    [I(t_i),H(t_i)] = [I(t_f),H(t_f)] = 0. 
\label{eq:bc}
\end{equation}
These are just necessary conditions, but they allow us to find a way to effectively speed up the adiabatic transfer between ground states at $t=t_i$ and $t=t_f$ in concrete cases, such as for atom transport \cite{Torrontegui2011}.

\textit{STA for atom transport.} A general family of Hamiltonians useful for atom transport and manipulation having a known dynamical invariant is given by \cite{Dhara1984} 
\begin{equation}
     H(t) = \frac{p^2}{2m} - F(t) x + \frac{m}{2}\omega^2(t)x^2
     + \frac{V\left(\frac{x-\alpha(t)}{\rho(t)}\right)}{\rho^2(t)}.
\label{eq:H_gen}
\end{equation}
It is possible to verify that for any constant $\omega_0$, the operator
\begin{align}
    I(t) = &\frac{\left[\rho(p - m\dot{\alpha})-m\dot{\rho}(x-\alpha)\right]^2}{2m} \\
           & + \frac{m\omega_0^2}{2}\left(\frac{x-\alpha}{\rho}\right)^2 +  V\left(\frac{x-\alpha}{\rho}\right)
\label{eq:I_gen}
\end{align}
is an invariant for $H(t)$ meaning that it satisfies Eq.~\eqref{eq:inv} provided that the following auxiliary conditions are also verified:
\begin{align}
    \ddot{\alpha} + \omega^2(t)\alpha &= \frac{F(t)}{m} \,, \text{ and }\label{eq:aux_alpha}\\
    \ddot{\rho} + \omega^2(t)\rho &= \frac{\omega_0^2}{\rho^3} \,. \label{eq:aux_rho}
\end{align}
As proven in Ref.~\cite{Dhara1984} there is a unitary transformation $\ket{\phi} = U(t)\ket{\psi}$, such that the transformed dynamical invariant $J = U I U^{\dagger}$, expressed in the coordinate $\zeta = \rho^{-1}(x-\alpha)$ is time independent. We can then obtain the eigenstates of $I$ by solving the stationary eigenvalue problem for $J$ and then transforming back with $U^{\dagger}$;
\begin{equation}
J\ket{\phi_n} =\lambda_n\ket{\phi_n}, \text{ with }  \ket{\psi_n(t)} = U^{\dagger}(t)\ket{\phi_n} \,,
\end{equation}
which in coordinate space give rise to \cite{Dhara1984,Torrontegui2011}
\begin{align}
\left[-\frac{\hbar^2}{2m}\frac{\partial^2}{\partial \zeta^2} + \frac{1}{2}m\omega_0^2\zeta^2 + V(\zeta)\right]\phi_n(\zeta) = \lambda_n \phi_n(\zeta), \\ 
\psi_n(x,t) = \rho^{-\frac{1}{2}}e^{\frac{im}{\hbar}[\dot{\rho} x^2/2\rho + (\dot{\alpha}\rho-\alpha\dot{\rho})x/\rho]}\phi_n\left(\frac{x-\alpha}{\rho}\right).
\end{align}
Finally, the Lewis-Riesenfeld phases are obtained as \cite{LR1969, Dhara1984}
\begin{equation}
    \gamma_n(t) = -\frac{1}{\hbar}\int_0^t{dt'}\left( \frac{\lambda_n}{\rho^2} + \frac{m(\dot{\alpha}\rho - \alpha\dot{\rho})^2}{2\rho^2} \right) ,
\end{equation}
and the boundary conditions from Eq.~\eqref{eq:bc}, which are necessary for eigenstate transfer, become
\begin{align}
    \dot{\alpha}(t_i) &= \ddot{\alpha}(t_i) = \dot{\alpha}(t_f) = \ddot{\alpha}(t_f) = 0, \\
    \dot{\rho}(t_i) &= \ddot{\rho}(t_i) = \dot{\rho}(t_f) = \ddot{\rho}(t_f) = 0 \,,
\end{align}
justifying Eqs.~\eqref{eq:bc_alpha} and \eqref{eq:bc_rho} in the main text.
By fixing $F(t)= m\omega^2(t)x_0(t), \text{ and } V(x)=0$ in Eq.~\eqref{eq:H_gen} we obtain the time-dependent harmonic oscillator Hamiltonian \cite{Torrontegui2011} from Eq.~ \eqref{eq:H_harm}, while Eqs.~\eqref{eq:aux_alpha}-\eqref{eq:aux_rho} become respectively Eqs.~\eqref{eq:q_0}-\eqref{eq:omga} of the main text.

\section{Time evolution method and discretization error analysis}
\label{sec_app_evol}
\renewcommand{\theequation}{B.\arabic{equation}}
\renewcommand{\thefigure}{B.\arabic{figure}}
\setcounter{figure}{0}

The numerical simulations inherently contain errors as an infinite Hilbert space is approximated by a finite one. For this reason examining the time evolution stability, as well as investigating the error due to space and time discretization becomes a crucial step in the validation of the obtained numerical results. With the aim of evaluating the numerical errors present in our model we study the Harmonic Oscillator (HO) problem in static and time-dependent conditions.

\textit{Space discretization error analysis.} The well-known solutions of the harmonic oscillator problem allow us to study the error introduced by the space discretization, constrained only by the machine precision. To that aim we first compare the probability for a wave function computed via exact diagonalization and the corresponding exact solution involving the Hermite polynomials for different space discretization steps $\Delta x$. The analysis offers an understanding of the error range in the considered interval for space discretization $\Delta x$. First we calculated the root mean square error between the two different computation methods when varying $\Delta x$. A fit of this quantity relates the error $\mathcal{E}$ to the discretization step $\Delta x$ as $ \mathcal{E}= a {\Delta x}^{b}/(1{+}c\Delta x)$ with $a=0.46$, $b=2.31$ and $c=7.03$, allowing for a selection of $\Delta x$ once the error threshold is fixed. 

Moreover, given that time required for the exact diagonalization algorithm scales as $\mathcal{O}(N^{3})$, with $N$ being the matrix dimension given by the size of the spatial grid \cite{demmel1997applied}, we are able to provide a rough estimation for the total computational time needed for our simulation for different $\Delta x$ to consequently approximate the computational time required to compute the Gaussian potential ground state. We found that $\Delta x \sim 0.02 \, \mu\text{m} \approx 0.2 l_{st} \approx 0.4 l_{mt}$ (where $l_{st}$ and $l_{mt}$ are respectively the characteristic lengths of static and moving trap during transport, i.e.\ with maximum tweezer depth) is sufficient to capture the wave-function details keeping a discretization error threshold of approximately $10^{-5}$ and a computational time in the range of milliseconds. 

\textit{Time discretization error analysis.} The time evolution is performed using the split-step Fourier method \cite{montangero2018introduction, lubich2008quantum, bao2002time, speth2013balanced, macnamara2016operator}, also beautifully explained in Ref.~\cite{hauck_prapp_2022}. This method is particularly effective thanks to the Baker-Campbell-Hausdorff relations \cite{cohen2019quantum}
\begin{equation}
\label{eq:BCH}
e^{i H\Delta t} \simeq e^{-i V \Delta t/2} e^{-i T \Delta t} e^{-i V \Delta t/2}  + \mathcal{O}(\Delta t^3) \, ,
\end{equation}
and the Fast Fourier Transform (FFT) algorithm \cite{frigo2005design}. Here $T$ and $V$ denote the kinetic and potential term of the Hamiltonian and we have chosen to reduce the error by introducing a symmetric splitting, also known as Strang splitting, where the potential is applied for half time step before and after the kinetic operator obtaining an error of $\mathcal{O}({\Delta t}^{3})$ \cite{macnamara2016operator,montangero2018introduction}. In practice, to take advantage of the kinetic operator being diagonal in the momentum space the Fourier transform is used in intermediate steps. This combination allows us to develop a straightforward algorithm to compute the evolution of the state $\ket{\psi(t)}$ from time $t$ to time $t+\Delta t$ that can be summarized in the following steps:
\begin{equation}
\begin{split}
e^{-i H \Delta t} \ket{\psi(t,x)} &\approx e^{-i V \Delta t/2} e^{-i T \Delta t} e^{-i V \Delta t/2} \ket{\psi(t,x)} \\
&\propto U_V(x) \mathcal{F}^{-1} \mathcal{F} U_T(x) \mathcal{F}^{-1} \mathcal{F} U_V(x) \ket{\psi(t,x)} \\
&\propto U_V(x) \mathcal{F}^{-1} U_T(p) \ket{\psi(t,p)} \\
&\propto U_V(x) \ket{\psi''(t,x)} \\
&\propto \ket{\psi(t+\Delta t,x)} \, ,
\end{split}
\end{equation}
where we have used $U_V = e^{-i V \Delta t/2}$, $ U_T = e^{-i T \Delta t}$, and $\mathcal{F}$ denotes the Fourier transform. The method assumes that the wave function does not change significantly within each small time step $\Delta t$. Also, from a computational perspective, the FFT scales as $\mathcal{O}(N \log(N))$, while the operators $U_V(x)$ and $U_T(x)$ appear in diagonal form (as mentioned before for the kinetic operatior to be diagonal the Fourier transform step is crucial), requiring storage of only $\mathcal{O}(N)$ elements and therefore speeding up the time evolution process. 

To study our system described by a Hamiltonian   
\begin{equation}
\label{Hamiltonian}
H(t) = \frac{p^2}{2m} + V(x) + V(x,t),
\end{equation}
we need to recursively apply the previous algorithm to the discretized potential $V(x,t_n) = V_n$ where $n \in [0, N_t]$ with $N_t$ being the total number of time steps, which in our computation for $T=3 \text{ ms}$ was chosen to be $5000$ (see below), leading to a $\Delta t = 0.6 \,\mu\text{s} \approx 0.02 \, \tau_{st} \approx 0.06 \, \tau_{mt}$. 

The FFT is the primary source of the high computational cost and a large number of time steps can further slow down the computation. Given the limited available computational resources, a compromise has to be made between the desired computational time and the required precision and stability for the simulation. One option for decreasing the computation time is to use the simple splitting method given by 
\begin{equation}
\label{eq:BCH_simple}
e^{i H \Delta t} \simeq e^{i V \Delta t}e^{i T \Delta t}  + \mathcal{O}(\Delta t^2) \, ,
\end{equation}
instead of Eq.~\eqref{eq:BCH}. It is important to highlight that the simple splitting leads to the ``correct'' results only at the initial and final steps, while introducing high-order oscillations for the intermediate steps. This can be understood if we explicitly write some of the steps in the recursive algorithm, 
\begin{equation}
\begin{split}
\ket{\psi(t_1)} &= e^{-i V_1\Delta t/2} e^{-i T \Delta t} e^{-i V_0\Delta t/2} \ket{\psi(t_0)} \\
\ket{\psi(t_2)} &= e^{-i V_2\Delta t/2} e^{-i T \Delta t} e^{-i V_1\Delta t/2} \ket{\psi(t_1)} \\
&\vdots \\
\ket{\psi(t_n)} &= e^{-i V_n\Delta t/2} e^{-i T \Delta t} e^{-i V_{n-1}\Delta t/2} \ket{\psi(t_{n-1})}
\end{split}
\end{equation}
where $\ket{\psi(t_0)},\ket{\psi(t_1)},...,\ket{\psi(t_n)}$ are the time evolved wave functions at times $t_0,t_1,...,t_n$.  At first sight by wrapping the consecutive potentials appearing in the recursive equation (for instance, while computing $\ket{\psi(t_2)}$ we need to use $\ket{\psi(t_1)}$ which at the end of the splitting contains $e^{-i V_1\Delta t/2}$, this can be wrapped with the first $e^{-i V_1\Delta t/2}$ of the computation for $\ket{\psi(t_2)}$) the Strang splitting methods seems to be computed in a similar way to the single splitting one. To quantify the error and the amplitude of the induced oscillations we studied the time evolution of a static harmonic oscillator by analyzing the evolution (for a fixed amount of $5000$ time steps) of the infidelity between the time-evolved state and the analytical one. We expect a vanishing infidelity given the static nature of the harmonic oscillator, however oscillations are observed for all the time steps different from the starting and the final one. The infidelity for the single splitting method oscillates around $1.7 \,10^{-4}$ while the infidelity obtained with the Strang splitting oscillates around $2.2 \,10^{-7}$ with the same oscillation period of two times the characteristic time of the oscillator. Even though the Strang splitting method is $30\%$ slower it leads to an improvement of more than two orders or magnitude in the infidelity which is crucial for us in order to get faithful results for infidelities under $10^{-4}$.

The time discretization $\Delta t$ was chosen by analyzing the splitting methods for different total number of time steps. After a careful quantification of the oscillations effects in the evolution methods we compute the infidelity between the numerical and analytical harmonic oscillator solution as a function of $\Delta t$ and chose a grid of $5000$ points for the larger considered total time of $3 \text{ ms}$ in order to reach infidelities below $10^{-7}$. With this procedure smaller times will keep the error bound. In all cases we observe that the Strang splitting improves the fidelity in at least 2 orders of magnitude. When varying the total evolution time and the frecuency of the oscillator in the interval of our interest we also checked that this two order difference remains, even though the error increases for increasing total times and frequencies. Finally, we used the trajectory obtained within our STA approach to move the considered harmonic oscillator and calculated the root mean square error between the numerically computed and the instantaneous analytical states. In the latter case we observe that the error for the Strang splitting is one order of magnitude less than the one obtained using the single splitting approach.  

\section{Transport dynamics}
\label{sec_app_transport_evol}

\renewcommand{\theequation}{C.\arabic{equation}}
\renewcommand{\thefigure}{C.\arabic{figure}}
\setcounter{figure}{0}

In this section we present the time evolution of the quantities studied in Sec.~\ref{sec_transport_char_opt} for some particular values of the total duration of the protocol denoted by $T$. We have chosen $T= 0.5, \, 1.5,\, 2.5,\, 3 \text{ ms}$ that corresponds to $16.6,\, 49.8,\, 83,\, 99.6\, \tau_{st}$ or $55.7,\, 167.1,\, 278.4,\, 334.1\, \tau_{mt}$. At this point it is important to remember that $\tau_{st}$ and $\tau_{mt}$ are the characteristic times of the static and moving trap during transport (i.e.\ with maximum tweezer depth) related by $\tau_{st} \approx 2\, \tau_{mt}$.

\begin{figure*}[t]
    \centering
    \includegraphics[width=\textwidth]{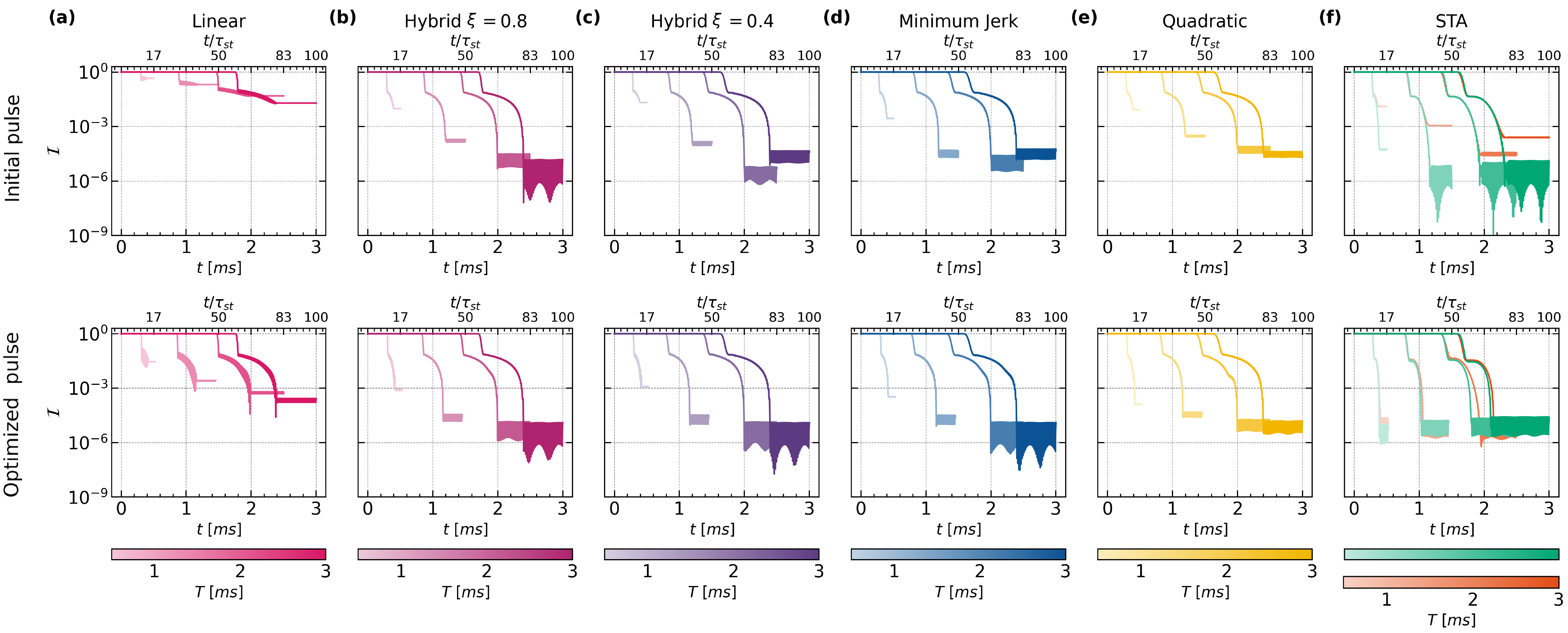}
    \caption{Error for transient times. Infidelity $\mathcal{I}$ vs. time for a total pulse time $T= 0.5, \, 1.5,\, 2.5,\, 3 \text{ ms}$ (lighter to darker color) for all the considered initial pulses (top row) and the corresponding optimizations (bottom). The time scales are given in SI and characteristic units, being $\tau_{st}$ the characteristic time of the static traps related to the one of the moving tweezer via $\tau_{st} \approx 2\, \tau_{mt}$. All the pulses have a maximum amplitude of $A_{mt}^{max}/\hbar=3.57 \times 2\pi \text{ MHz}$, the transport time is given by $\eta \, T$ with $\eta=2/5$, and the total covered distance between static tweezers is $d = 7\, \mu\text{m}$.}
    \label{fig_inf_evol}
\end{figure*}

In figure \ref{fig_inf_evol} we report the error after transport measured as the maximum of the infidelity with respect to the target state over the last interval of the pulse [see Eqs.~\eqref{A_mt_of_t} and \eqref{x_mt_of_t}] for the five families of pulses we consider in the main text. The upper row shows the results without optimization while the error during transport obtained for the optimized pulses is depicted in the bottom row. The darker the color of the curve, the larger the total time $T$. As can be seen, the evolution of the infidelity depicts more or less the same behavior for all the pulses; the fidelity is very low until the atom reaches the target tweezer, after that the fidelity rapidly increases. During the last stage of the pulse (waiting time) the value of the infidelity presents some oscillations that can be seen as the thick final part of each curve. To interpret this results it is important to have in mind that the data are presented in logarithmic scale, therefore, even though the STA final evolution seems to be very noisy, the final variations are spread on a $10^{-6}$ scale. Also, for the hybrid pulse with $\upxi = 0.4$ and the minimum jerk ramp we see that the final infidelity is higher for $T=3 \text{ ms}$ than for $T=2.5 \text{ ms}$, this is consistent with the oscillation that the infidelity presents when plotted against $T$ (see Fig. \ref{fig:ts_03} in the main text). All the optimized pulses improve the fidelity in at least one order of magnitude.

\begin{figure*}[t]
    \centering
    \includegraphics[width=\textwidth]{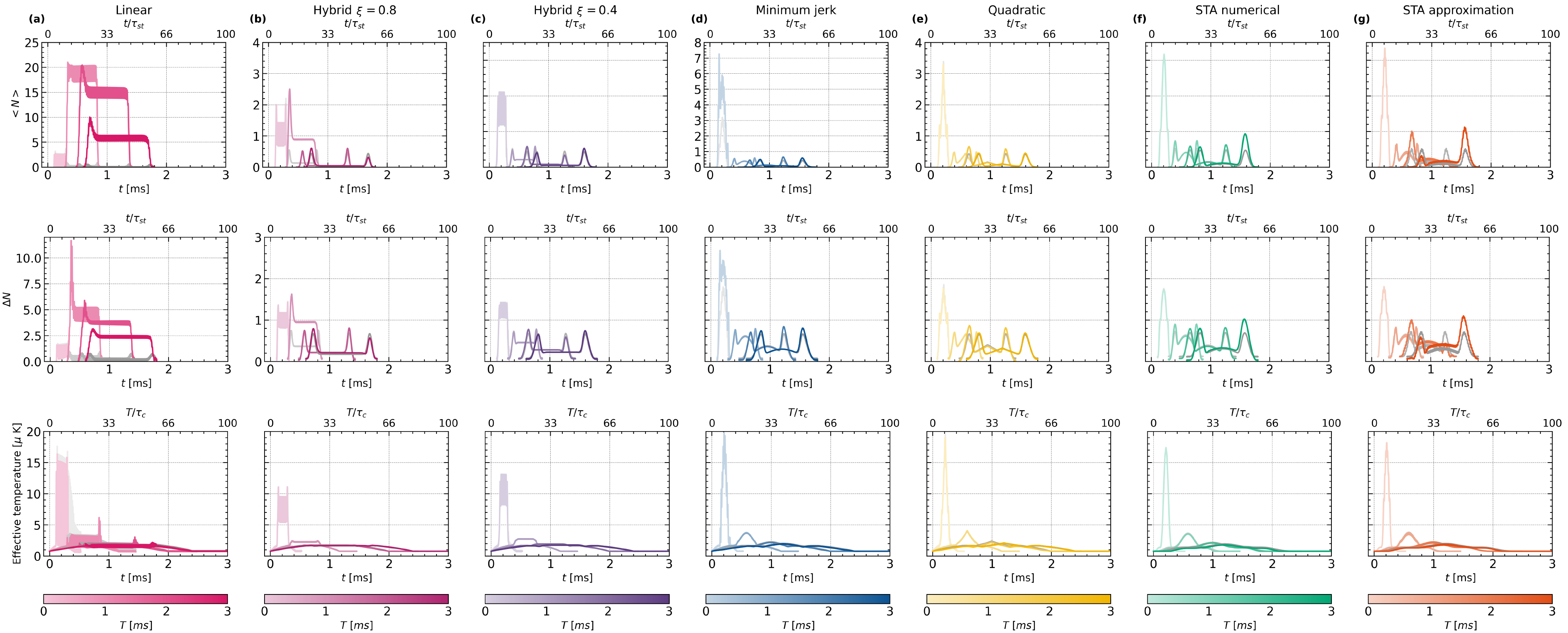}
    \caption{Measures of heating for transient times. Expected value $\expval{N}$ for the occupied level of the moving trap states (top row) with the associated uncertainty $\Delta N$ (middle row), and effective temperature $T_\text{eff}$ (bottom row) vs. time for a total pulse time $T= 0.5, \, 1.5,\, 2.5,\, 3 \text{ ms}$ (lighter to darker color) and for all the considered pulses. The color curves correspond to the optimized pulses, the results obtained with the initial (non optimized) pulses are very similar (except for the piece-wise linear pulse) and therefore are shown in gray behind each optimized curve. The time scales are given in SI and characteristic units, being $\tau_{st}$ the characteristic time of the static traps related to the one of the moving tweezer via $\tau_{st} \approx 2\, \tau_{mt}$. All the pulses have a maximum amplitude of $A_{mt}^{max}/\hbar=3.57 \times 2\pi \text{ MHz}$, the transport time is given by $\eta \, T$ with $\eta=2/5$, and the total covered distance between static tweezers is $d = 7\, \mu\text{m}$.}
    \label{fig_dN_Teff_of_t}
\end{figure*}

Figure \ref{fig_dN_Teff_of_t} shows the time evolution of the mean occupied level of the moving trap states $\expval{N}$ and its uncertainty $\Delta N$ given by the width of the distribution over the states of the moving tweezer during transport (upper row) as well as the motional effective temperature $T_\text{eff}$ (lower row). Since the non-optimized and optimized curves are very similar except for the linear pulse, the values for the initial (non optimized) ramps are shown as gray curves and the values obtained for the optimized pulses are in color. As before darker colors indicate larger total times. Shorter pulse times are related to higher $\expval{N}$, $\Delta N$, and $T_\text{eff}$. Also, the optimization induces heating mainly for the piece-wise linear ramp, the hybrid one with hybridicity $\upxi =0.8$ (more close to the linear pulse), and the STA approximation. We notice that the distribution over the moving tweezer states is more sensitive as heating measure for transient times than the effective temperature. The two peaks depicted by $\expval{N}$ and $\Delta N$ correspond to the mixing of states induced when the wave function abandons one of the tweezers and readjusts into the following one. As expected due to the jumps in the velocity, in general, the piece-wise linear ramp depicts the higher $\expval{N}$ and $\Delta N$.  

\begin{figure*}[t]
    \centering
    \includegraphics[width=0.95\textwidth]{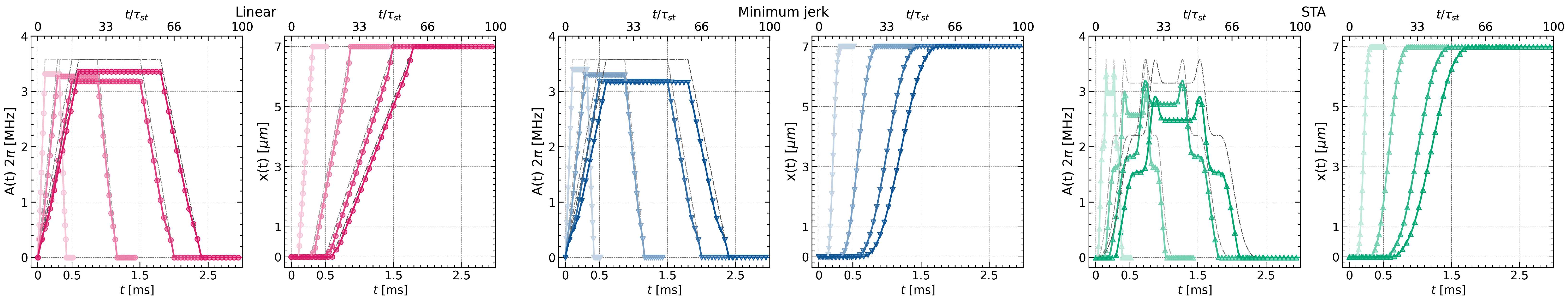}  
    \caption{Examples of optimized pulses. Amplitude $A_{mt}(t)$ and position $x_{mt}(t)$ of the moving tweezer for the linear (left), minimum jerk (center), and STA (right) ramp. The optimized pulses are shown in color solid curves while the initial (non optimized) ones are depicted as gray dashed lines. In general we observe that the optimization performs small changes in the minimum jerk and STA trajectory compared to the rest of the pulses. As before, the time scales are given in SI and characteristic units, being $\tau_{st}$ the characteristic time of the static traps related to the one of the moving tweezer via $\tau_{st} \approx 2\, \tau_{mt}$. All the pulses have a maximum amplitude of $A_{mt}^{max}/\hbar=3.57 \times 2\pi\text{ MHz}$, the transport time is given by $\eta \, T$ with $\eta=2/5$, and the total covered distance between static tweezers is $d = 7\, \mu\text{m}$.}
    \label{fig_opt_pulses}
\end{figure*}

Finally, figure \ref{fig_opt_pulses} presents examples of the comparison between the optimized pulses (color solid lines) and the initial guesses or non optimized pulses (gray dashed lines) for the same total times considered before $T= 0.5, \, 1.5,\, 2.5,\, 3 \text{ ms}$. We would like to highlight that the optimized position pulse for the linear ramp is not symmetric. This can also be seen in the asymmetric form depicted by $\expval{N}$ and $\Delta N$ in figure \ref{fig_dN_Teff_of_t}. While all the considered initial pulses were taken symmetric, the automatic optimization algorithm breaks this symmetry specially for the piece-wise linear and approximated STA pulse. Also, we observe that the optimization changes the piece-wise linear pulse mainly at the beginning and end of the transport stage, also for the STA pulse the optimization tends towards smaller amplitudes.

\bibliographystyle{myieeetr}

\end{document}